\documentclass{aa}  

\usepackage{graphicx}
\usepackage{txfonts}
\usepackage{xcolor}
\begin{document} 

   \title{The wide binary frequency of metal-poor stars}
%
%
   \author{N.\ Lodieu \inst{1,2}
       \and
       A.\ P\'erez Garrido \inst{3}
       \and
       J.-Y.\ Zhang \inst{1,2}
       \and
       E. L.\ Mart\'in \inst{1,2}
       \and
       R.\ Rebolo L\'opez \inst{1,2,4}
       \and
       F.\ P\'erez-Toledo \inst{5}
       \and 
       \newline
       R.\ Clavero \inst{1,2}
       \and
       D.\ Nespral \inst{1,2}
        }

    \institute{Instituto de Astrof\'isica de Canarias (IAC), Calle V\'ia L\'actea s/n, E-38200 La Laguna, Tenerife, Spain \\
    \email{nlodieu@iac.es}
    \and
    Departamento de Astrof\'isica, Universidad de La Laguna (ULL), E-38206 La Laguna, Tenerife, Spain
    \and
    Departamento de F\'isica Aplicada, Universidad Polit\'ecnica de Cartagena, Campus Muralla del Mar, 30202 Cartagena, Murcia, Spain
    \and
    Consejo Superior de Investigaciones Cient\'ificas (CSIC), E-28006 Madrid, Spain
    \and
    GRANTECAN, Cuesta de San Jos\'e s/n, 38712 Bre\~na Baja, La Palma, Spain
       }

   \date{\today{},\today{}}

 
  \abstract
   {This study is aimed at identifying possible low-mass and sub-stellar companions to stars with well-determined metallicities. We investigate the multiplicity of metal-poor stars along with its impact on formation processes in the conditions of the early universe.
   }
   {Our goal is to look for wide common proper motion companions to metal-poor stars and study the binarity frequency at low metallicity with astrometry from large-scale catalogues, including $Gaia$, Visible and Infrared Survey Telescope for Astronomy (VISTA) Hemisphere Survey (VHS), and Wide Field Infrared Survey Explorer (WISE).
   }
   {We used the stellar parameter determination from the latest release of the $Gaia$ catalogue to identify metal-poor stars over the entire sky. We combined the $Gaia$ sample with other public catalogues and spectroscopic determinations for a given sub-sample to refine the stellar metallicities. We also considered, as input, other public catalogues of metal-poor stars to look for co-moving companions. We also obtained our own high-resolution images of a sub-sample with the lucky imaging technique.
   }
   {We only found a few bona fide co-moving systems among a sample of 610 metal-poor stars with metallicities below $-$1.5 dex in the full sky. We inferred a multiplicity rate below 3\%, with 3$\sigma$ completeness for projected separations larger than 8 au, after taking into account incompleteness and any other limiting factors of our search. At closer separations, we found a minimum binary fraction of 20\% that appears to be relatively independent of metallicity.
   }
   {We conclude that the multiplicity fraction of solar-type stars is relatively independent of metallicity for close-in companions with projected separations below $\sim$8 au. At separations between 8 and 10000 au, the binary fraction of metal-poor stars drops significantly to a few percent and is significantly lower than the multiplicity derived for the solar-metallicity case. We interpret these similarities and differences as being due to the chemistry at work in molecular clouds as well as disruption effects attributed to the old age of sub-dwarfs.}

   \keywords{surveys -- astrometry -- metal-poor stars -- binaries}
               
\maketitle

%
%
\section{Introduction}
\label{CPM_MP:intro}

The multiplicity of stars is essential in several areas of astronomy, such as stellar evolution, dynamics, and gravitational theories. Binary stars provide critical information that can be used to test and refine theoretical stellar models \citep{popper80,andersen91a,southworth05,torres10a}.
The study of multiple systems sheds light on star formation processes and the dynamical effects, bringing on new insights into star formation and galactic evolution through their evolution over time. Binaries are key systems for measuring stellar masses accurately, even including radii, in cases of eclipsing binaries when applying Kepler's third law. This information is fundamental to infer the stellar properties of each component and provides key inputs to theoretical models of stellar structure and evolution \citep{lodieu20a}. The observational determination of masses, temperatures, and luminosities of stars allow for the validation or refining of existing models. 

Metal-poor stars, also known as sub-dwarfs, are astronomical objects with abundances of elements heavier than hydrogen and helium in their atmospheres at levels that are at least ten times lower compared to the Sun and other Population I stars in the Milky Way. These stars have ages around 10--13 billion years, making them among the oldest objects in the Universe, having formed early on in the history of the Milky Way. They exhibit high proper motions and heliocentric velocities with kinematic properties typical of the thick disk or halo of the Galaxy with velocities typical larger than 60--100 km/s \citep{gizis97a,lepine07c,jao08}. Subdwarfs typically exhibit lower luminosities than their solar counterparts and higher gravities, hence, smaller radii as well, compared to giants and subgiants. Metal-poor dwarfs usually bluer colours mainly due to enhanced calcium hydrides (CaH) in the optical and stronger collision-induced absorption-affected near-infrared (NIR) wavelengths. Sub-dwarfs provide valuable information about the conditions characterising the early universe and the processes that led to the formation of the first generation of stars in the Milky Way and other galaxies \citep{buser00a,bland_hawthorn16,gallart19}.

Stellar multiplicity is the result of a complex interplay of many physical processes, including metallicity, initial formation conditions, stellar evolution, and environment. Metal-poor environments tend to have lower opacities and cooling rates, which can affect the fragmentation process during star formation. Lower metallicity gas is less efficient at radiating away heat, leading to the formation of fewer but more massive fragments \citep{bate19a,myers14a,chon21b,guszejnov22b,bate23}. Comprehensive studies of the multiplicity of metal-poor stars, including observational surveys, statistical analyses, and theoretical insights can be found in several key papers and references therein; for instance \citet{abt83}, \citet{carney83a}, \citet{carney83b}, \citet{latham02}, \citet{raghavan10}, \citet{moe17}, \citet{gao17a}, \citet{badenes18a}, \citet{El_Badry19a}, \citet{liu19a}, \citet{mazzola20a}, \citet{hwang21a}, and \citet{niu22a}.

Several surveys have specifically looked into the close binary fraction of stars at different metallicity, with some controversial outcomes \citep[e.g.][]{stryker85,carney94,latham02}, as well as differences settled in the work by \citet*{moe18b} where the metallicity does play a role in the frequency of close-in binaries with projected separations below 10 au. These authors showed that stellar interactions are an important factor that links binary frequency with metallicity. The overall frequency of F- and G-type sub-dwarfs \citep[30--40\%;][]{latham02,goldberg02a,carney03,jao09a,moe17,El_Badry19a} seems slightly lower than for their solar-metallicity counterparts \citep[50--60\%][]{duquennoy91,raghavan10}. Similarly, the global frequency of M sub-dwarf binaries is lower than their solar-metallicity counterparts \citep[15\% vs 30--35\%;][]{jao09a,riaz08a} in addition to a lower number of wide co-moving pairs \citep{riaz08a,jao09a,lodieu09c,ziegler15}; this is a trend that was previously observed for FGK sub-dwarfs as well \citep{hwang21a,niu22a}. The separation between components of M sub-dwarf multiples is lower and the mass ratio closer to unity than at solar-metallicity \citep{fischer92,reid97a,delfosse04,dhital10,bergfors10}.

This paper presents a preliminary study of the multiplicity of metal-poor stars with metallicities below $-$1.5 dex over a large range of projected physical separations accessible to $Gaia$ and seeing-limited images.
In Section \ref{CPM_MP:sample} we describe the sample extracted from the $Gaia$ latest data release. In Section \ref{CPM_MP:search} we present several independent searches for wide companions: one based on $Gaia$ data only, another one cross-matching the VISTA Hemisphere Survey \citep[VHS;][]{mcmahon12} and Wide Field Infrared Survey Explorer \citep[AllWISE;][]{wright10}, along with a lucky imaging survey of a sub-sample of stars.
In Section \ref{CPM_MP:multiplicity} we put the numbers of common proper motion pairs identified in our search into context to determine the multiplicity of metal-poor stars and make comparisons with previous studies in the literature.

%
%
%
\begin{figure*}
   \centering
   \includegraphics[width=0.32\linewidth, angle=0]{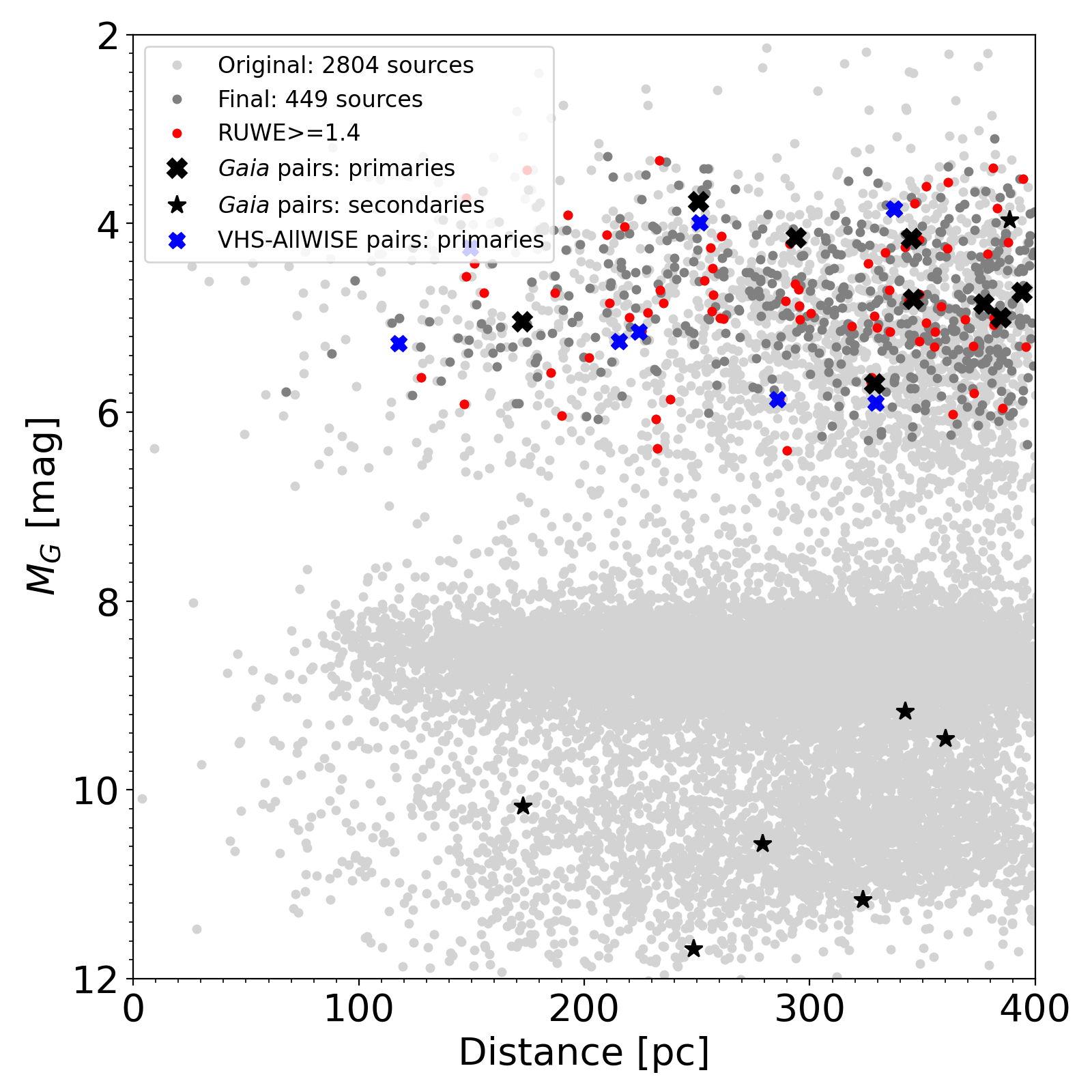}
   \includegraphics[width=0.32\linewidth, angle=0]{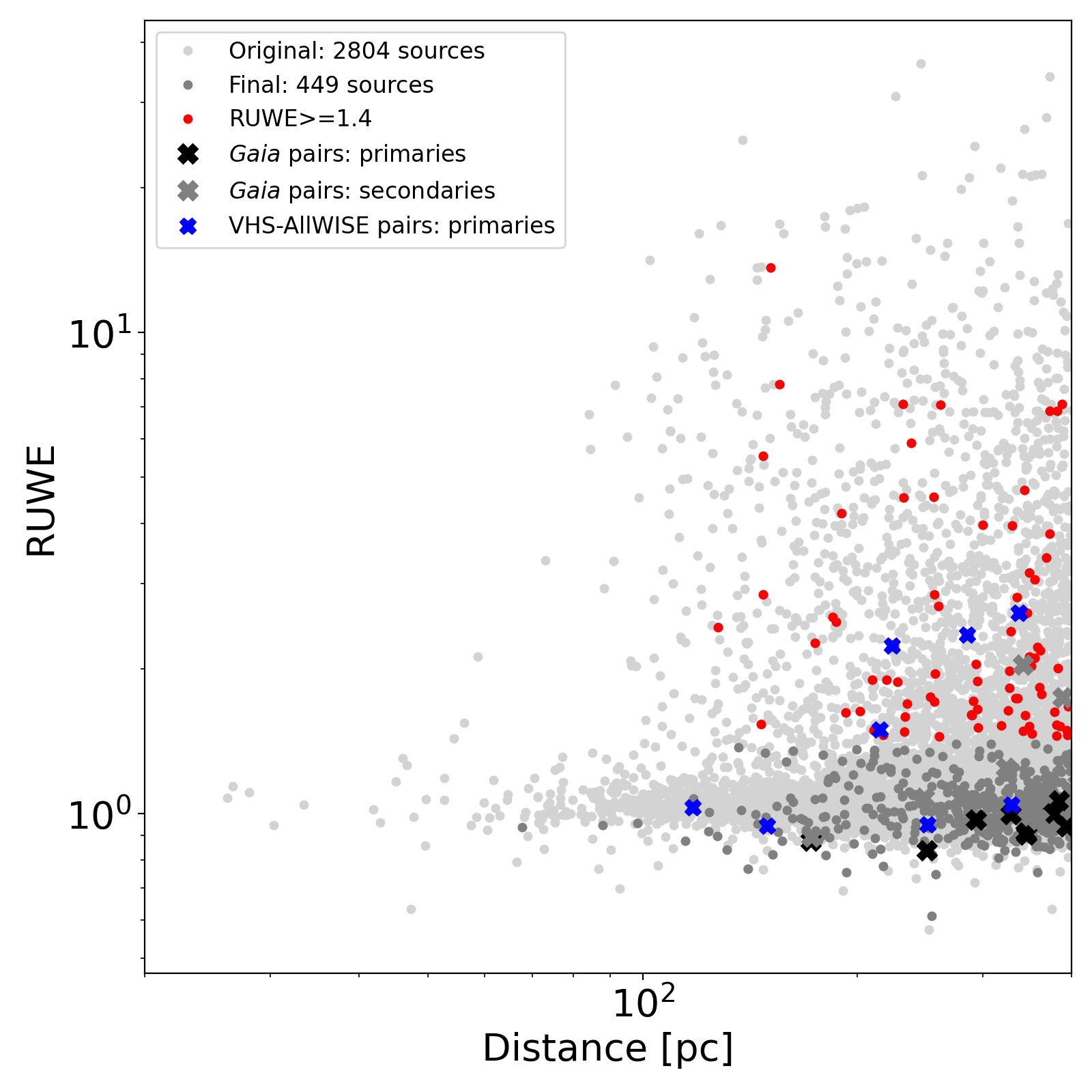}
   \includegraphics[width=0.32\linewidth, angle=0]{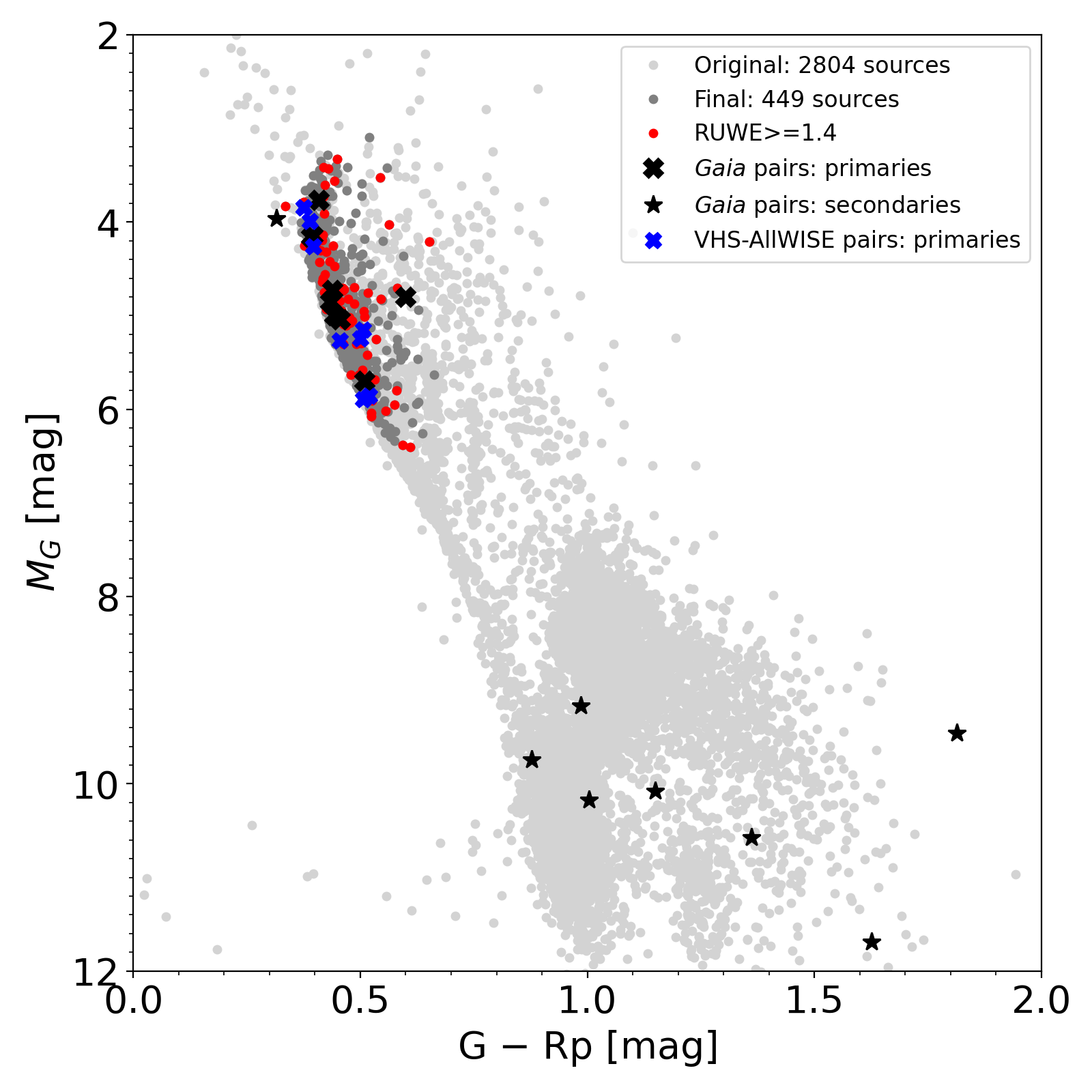}
   \includegraphics[width=0.32\linewidth, angle=0]{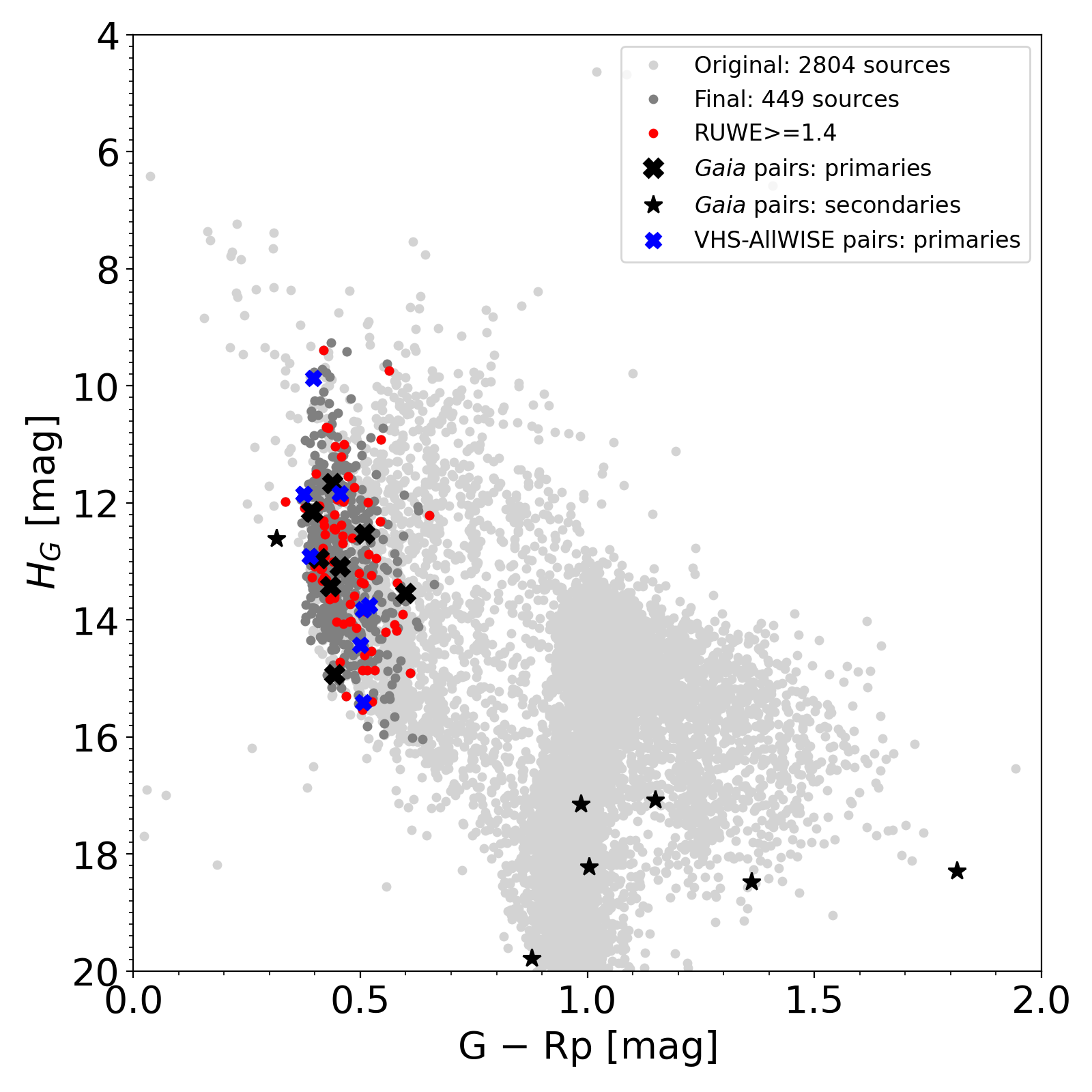}
   \includegraphics[width=0.32\linewidth, angle=0]{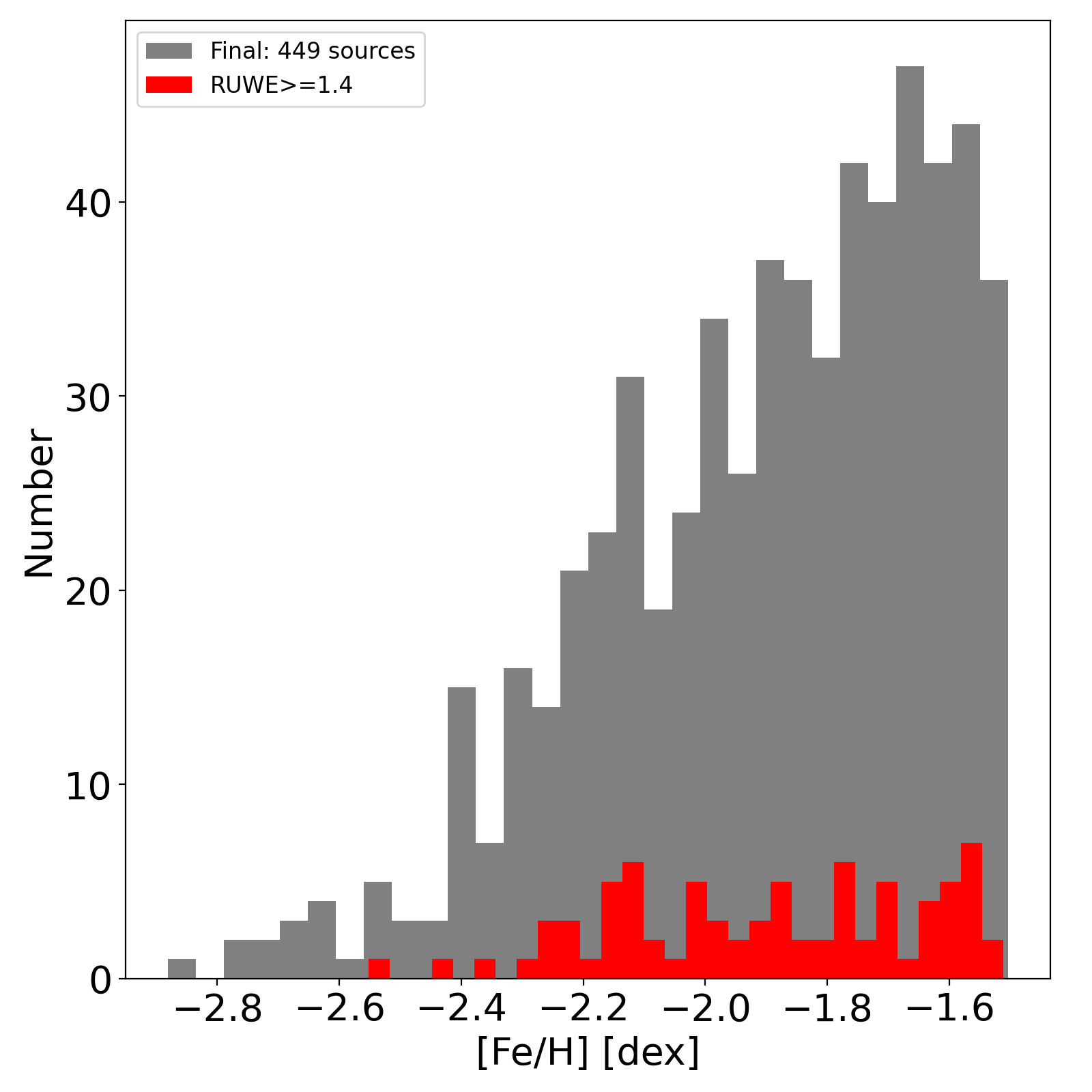}
   \includegraphics[width=0.32\linewidth, angle=0]{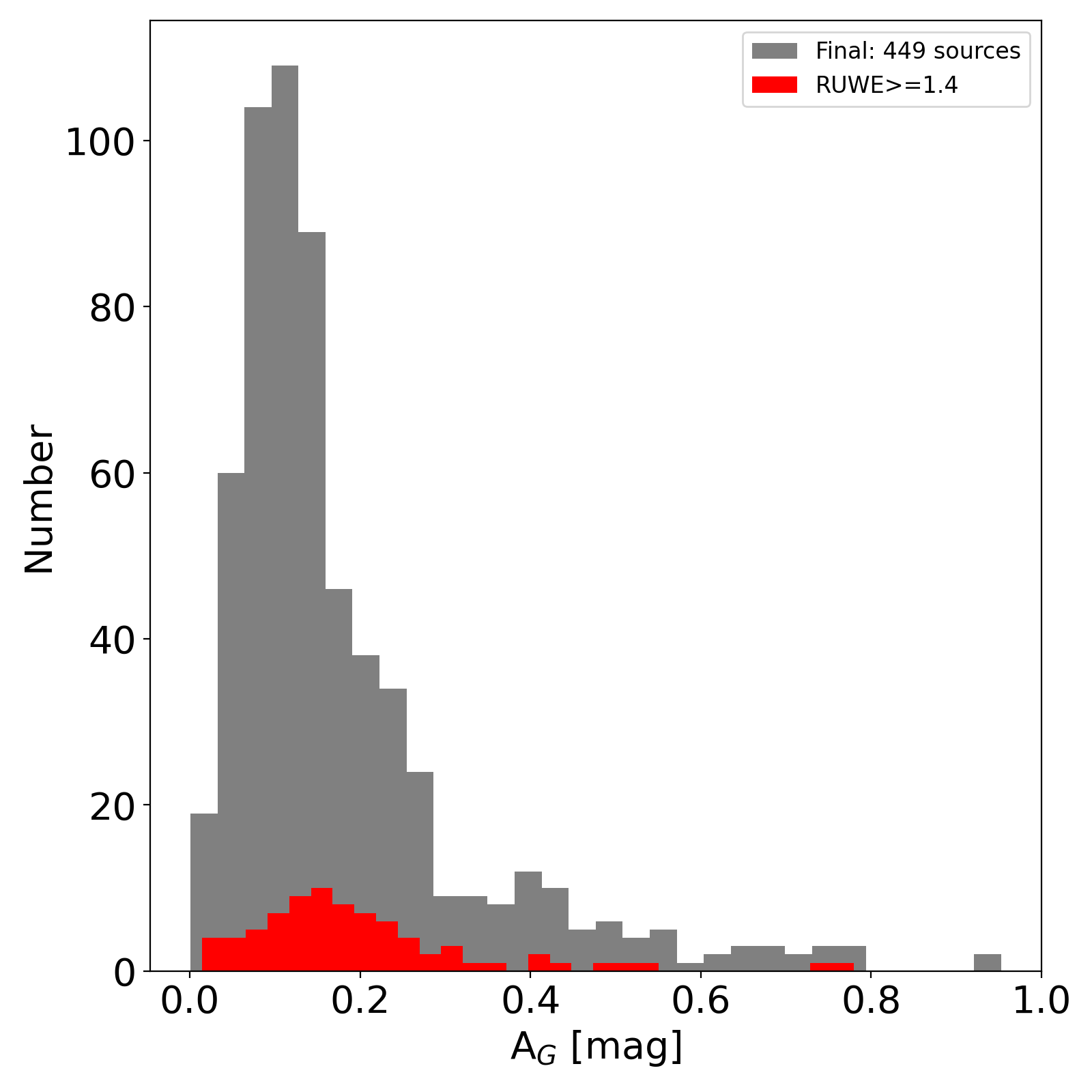}
   \caption{Main parameters of the original (light gray) and final (gray) samples. Sources with high RUWE (greater or equal to 1.4) are highlighted in red. Multiple systems identified in the $Gaia$ and VHS-AllWISE catalogues are marked with large black and blue symbols, respectively.
   }
   \label{fig_CPM_MP:sample}
\end{figure*}
%

%
%
\section{Sample selection}
\label{CPM_MP:sample}

We launched an Astronomical Data Query Language (ADQL; Appendix \ref{CPM_MP:ADQLquery}) query in the $Gaia$ science archive\footnote{\url{https://gea.esac.esa.int/archive/}} to retrieve all metal-poor stars in the $Gaia$ Data Release 3 \citep[DR3;][]{gaia_collaboration2016,vallenari23}. We set the following constraints to identify good quality sources with physical parameters derived by $Gaia$ \citep{creevey23,fouesneau23}, namely: distances below 400 pc with 3$\sigma$ detections, metallicities below $-$1.5 dex, and transversal velocities higher than 60 km/s \citep{tian20a}. The choice of the upper limit on the metallicity is our own choice to look at truly metal-poor dwarfs with little contamination from solar-metallicity sources due to the current uncertainties in the determination of metallicities that can be up to 1 dex. The query returned 15432 sources (hereafter ``our sample''), which can be easily downloaded from the $Gaia$ archive by running the ADQL query provided in Appendix \ref{CPM_MP:ADQLquery}. 

\begin{figure}
 \centering
 \includegraphics[width=\linewidth, angle=0]{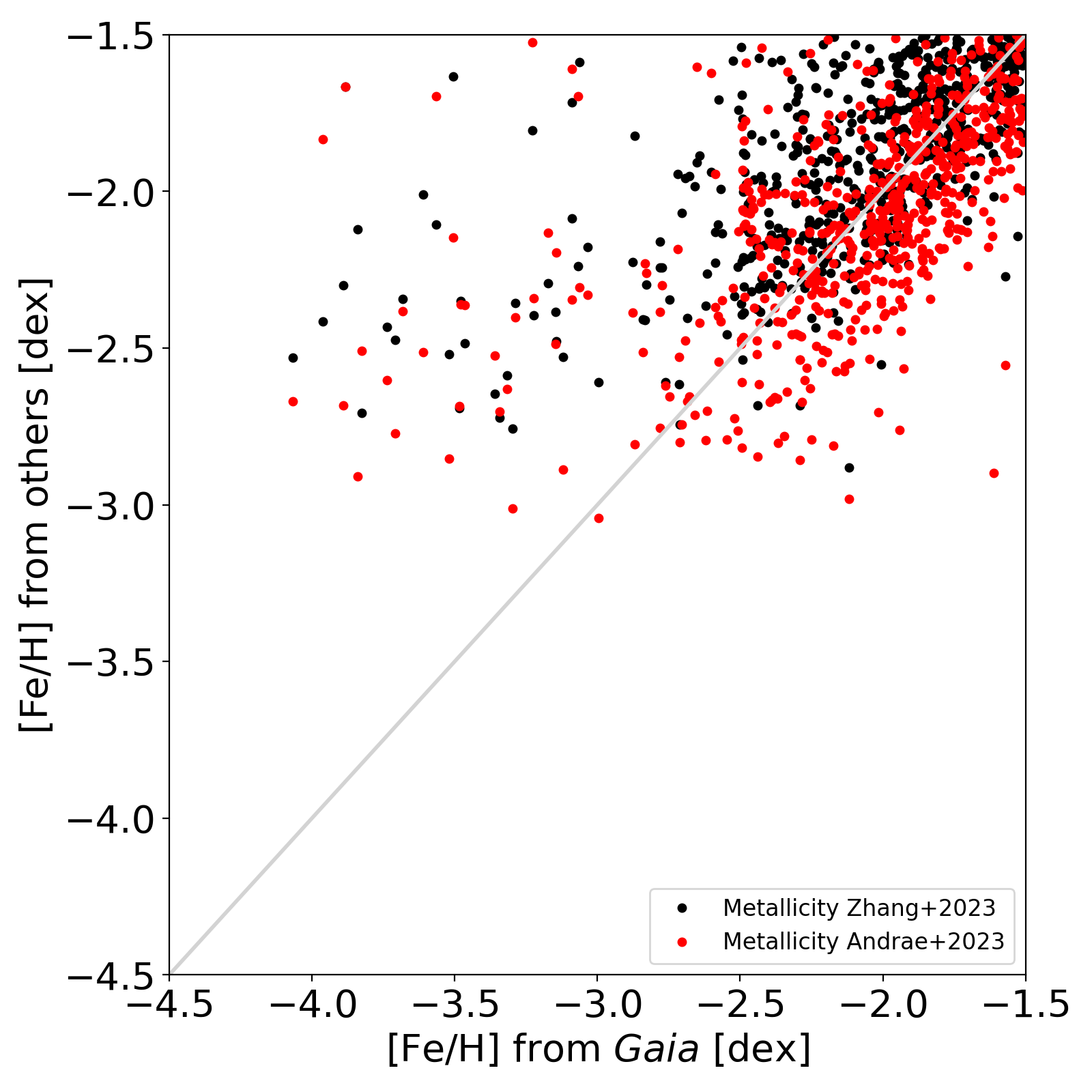}
 \caption{
 Comparison of the $Gaia$ metallicities (x-axis) with the determinations from \citet{andrae23b} in red and from \citet{zhang23a} in black.
 }
  \label{fig_CPM_MP:FeH_compare_main}
\end{figure}

The $Gaia$ metallicities suffer from shortcomings (e.g.\ systematics and outliers) that other teams have looked into to derive metallicities with higher fidelity and fewer outliers (left-hand side plot in Fig.\ \ref{fig_CPM_MP:FeH_compare_main}), particularly below $-$2.5 dex. As a consequence, we have cross-matched our sample with the catalogues published by \citet{andrae23b}\footnote{\url{https://zenodo.org/records/7945154}} and \citet{zhang23a}\footnote{\url{https://zenodo.org/records/7811871}}. We note that we imposed a constraint on the quality flag in this catalogue: {\tt{quality\_flags}}\,$\leq$\,8 that contain good quality 175 and 220 million sources with $Gaia$ spectra and more reliable metallicities. If we restrict these catalogues to the parameters set for our sample, we end up with 6500 and 1589 metal-poor ([Fe/H]\,$\leq$\,$-$1.5 dex) sources with parallaxes larger than 2.5 mas with a 3$\sigma$ confidence. After cross-matching our sample with both catalogues using a matching radius of 3 arcseconds (arcsecs), we identified 610 sources in common that constitute our main best-confidence sample of study in the rest of this work. The main difference with the original catalogue is the disappearance of the sources with high extinction (Fig.\ \ref{fig_CPM_MP:sample}.)
The key properties of the samples are plotted in Fig.\ \ref{fig_CPM_MP:sample}. Most of the stars in this sample have effective temperatures and gravities in the range 5000--6500\,K \citep[approximately mid-F to early-K;][]{pecaut13} and 3.8--4.6 dex, respectively \citep{andrae23b}.


We also added sources from other catalogues of metal-poor stars freely available in the literature. In particular, we considered the Jinabase database of metal-poor stars \citep{abohalima18}\footnote{\url{https://jinabase.pythonanywhere.com}} as well as the sample of 385 very metal-poor stars, including the spectroscopic follow-up \citep{li22a}. To ensure a reliable identification of common proper motion candidates in both samples, we limited our search to objects with total $Gaia$ proper motions larger or equal to 200 mas/yr. Only half of the stars in the Jinabase catalogue satisfy this astrometric criterion, while just ten stars remain in the sample of \citet{li22a}, most of them outside the VHS coverage. We kept both samples because of the trustworthy physical parameters and low metallicities, but we caution that the statistics on both catalogues is extremely limited.

%
%
%

%
%
\section{Search for wide faint companions}
\label{CPM_MP:search}
\subsection{Search for common proper motion companions}
\label{CPM_MP:search_Gaia}

We cross-correlated internally our main sample of 610 sources varying the matching radius up to 300 arcsecs. This search did not returned any wide pair. $Gaia$ is expected to be complete for binaries with separation of a few arcsecs down to its full depth. However, our cut in metallicity is strict at $-$1.5 dex and some true common proper motion candidates to sources in the main sample may exist with slightly higher metallicities or no [Fe/H] measurement at all. 

We cross-matched these 610 sources with $Gaia$ DR3 up to a separation of 25 arcsecs, yielding 4309 candidates. After removing the primary sources themselves and keeping only common proper motion candidates at similar distances within 3$\sigma$ of the error bars in parallax, $\mu_{\alpha}\cos\delta$, and $\mu_{\delta}$, we are left with nine pairs (Table \ref{tab_CPM_MP:tab_wide_comp} in Appendix \ref{CPM_MP:wide_pairs_300arcsec}; large black symbols in Fig.\ \ref{fig_CPM_MP:sample}).
The $Gaia$ metallicity of the secondaries are either above the $-$1.5 dex limit or absent from the $Gaia$ DR3 catalogue; this indicated the reason why they escaped the original selection. None of the primaries have large  renormalised unit weight error \citep[RUWE;][]{Gaia_Lindegren2018}, whereas two secondaries have values larger than 1.25, suggesting that they might be binaries with separations below 0.02 arcsec that are unresolved by $Gaia$ \citep{penoyre20,belokurov20a,gandhi22a}. The chance alignment is extremely small at these separations is displayed in Fig.\ 4 of \citet{El_Badry21a}. Therefore, these systems might actually be higher order multiple. These systems seem to suffer little extinction (A$_{G}$) as computed by $Gaia$ in the $G$-band along the line-of-sight (histogram at the bottom-right panel in Fig.\ \ref{fig_CPM_MP:sample}).

We found an optical spectrum of the source 00:49:20.33 $+$06:25:09.7 in the latest data releases of LAMOST\footnote{\url{http://www.lamost.org/dr10/v2.0/}}. The spectral type derived by LAMOST is A9--F0 with a temperature of 6075--6133$\pm$34\,K, a metallicity between $-$1.85$\pm$0.01 and $-$1.77$\pm$0.02 dex and a large radial velocity of about $-$188$\pm$5 km/s. We found an APOGEE spectrum of 14:53:10.23$+$48:14:43.8 in the latest release of the SDSS-III\footnote{https://skyserver.sdss.org/dr18/} survey. The inferred parameters are T$_{\rm eff}$\,=\,5759$\pm$59\,K, [Fe/H]\,=\,$-$1.83$\pm$0.02 dex, and RV\,=\,$-$156.8$\pm$0.2 km/s. Thus, the primaries of both systems have been spectroscopically confirmed as metal-poor dwarfs with metallicities in agreement with the ones in the $Gaia$ catalogue to better than 0.3 dex.

\begin{figure}
 \centering
 \includegraphics[width=\linewidth, angle=0]{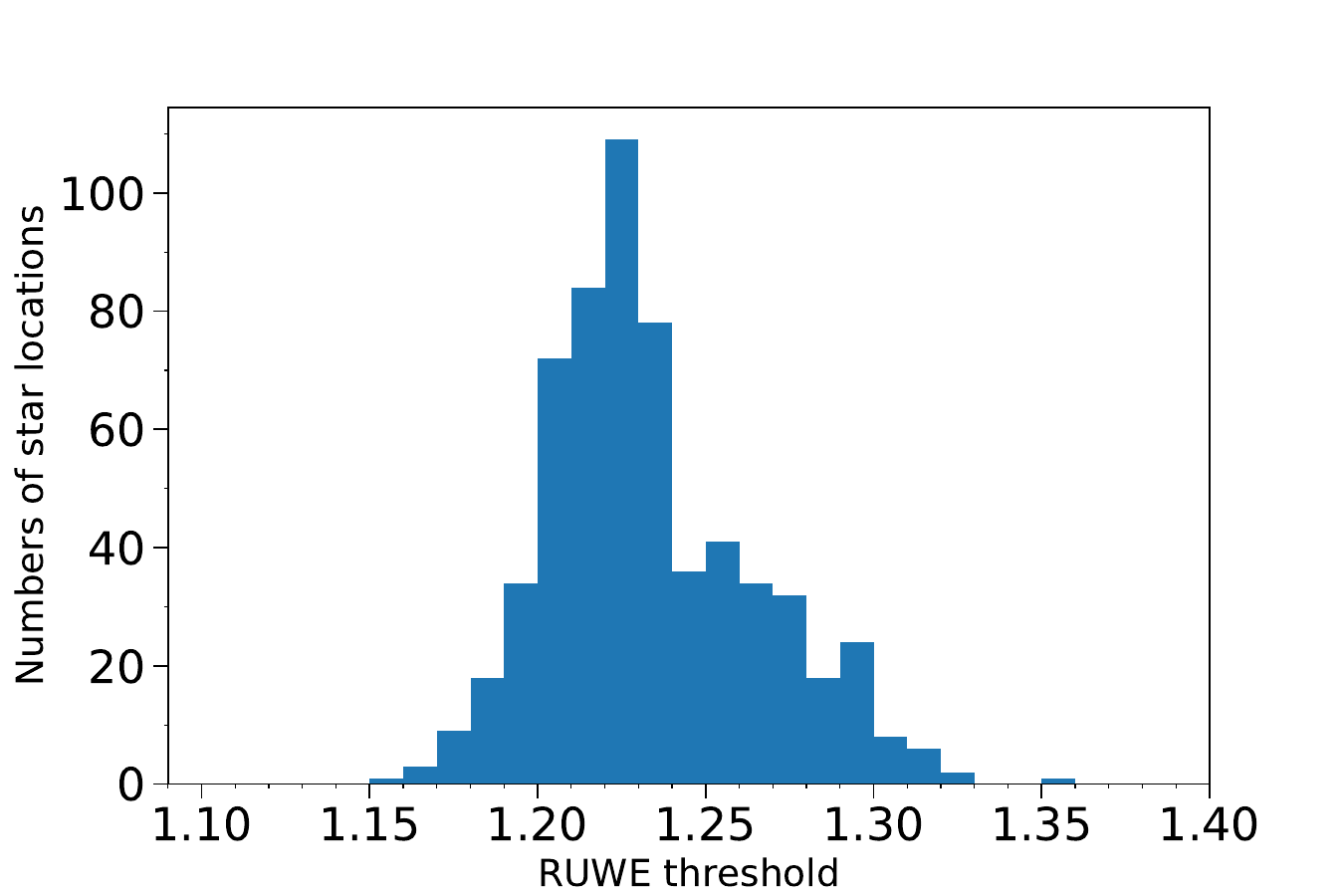}
 \caption{Histogram of the RUWE parameter across the sky for our sample of sub-dwarfs following the simulation from \citet{castro_ginard24}.
 }
  \label{fig_CPM_MP:histogram_RUWE}
\end{figure}
\subsection{Search for faint wide companions}
\label{CPM_MP:search_VHS_AllWISE}

The VISTA Hemisphere Survey \citep[VHS;][]{mcmahon12} is a NIR survey of
the entire Southern Sky (20,000 square degrees) using the 4m Visible and Infrared Survey Telescope for Astronomy \citep[VISTA;][]{emerson01,emerson04} equipped with a camera made of sixteen 2048$\times$2048 pixel detectors covering 1.65 square degrees, after dithering, with a mean pixel scale of 0.34 arcsec \citep{dalton06}. Half of the total area is covered in $J$ and $K_{s}$ while 5,000 deg$^{2}$ are covered in $JHK_{s}$ and
another 5,000 deg$^{2}$ in $YJHK_{s}$.
The mean 5$\sigma$ point source depth of the combined images are about $Y$\,=\,20.0 mag, 
$J$\,=\,19.5 mag, $H$\,=\,18.6 mag, and $K_{s}$\,=\,17.9 mag in the Vega system.

The All-Sky Wide-Field Infrared Survey Explorer (AllWISE) space-based mission covered the entire sky in four mid-infrared (MIR) passbands: 3.4 ($W1$), 4.6 ($W2$), 12 ($W3$), and 22 ($W4$) microns down to 17.1, 15.7, 11.7, 7.7 mag (95\% completeness), respectively \citep{wright10}. It was a follow-up to the WISE mission, which operated from 2009 to 2011, with improved data processing and coverage. The total number of entries in the AllWISE source catalogue with $W1$ detections is close to 750 million sources, number decreasing with longer wavelength.

We cross-matched VHS and AllWISE to look for companions to metal-poor stars in the final common sample selection. We find $\sim$44\% of these objects (268/610) covered by VHS\@. We prepared lists of objects from both catalogues that are located at a distance of less than 180 arcsecs from each source in the selection, discarding objects that do not show motion signatures between VHS and AllWISE\@. We remained sensitive to companions about 1.5 times the full width at half maximum (FWHM) of WISE in the $W1$ and $W2$, which is about 10 arcsecs. Then, we prepared lists of pair of sources with a separation between VHS and AllWISE less than 60 arcsecs. This places an upper limit in proper motion of about 10 arcsecs/yr. 
We computed the proper motion for each pair in the lists rejecting pairs with a difference in proper motion with the object in the $Gaia$ sample larger than a 15\%. The output lists contains mostly primaries from the $Gaia$ sample along with contaminants and background extended objects, which were discarded after visual inspection. This search returned eight potential pairs with separations larger than 80 arcsecs and $J$-band magnitude in the 17.6--20.3 mag range after excluding the primaries themselves, placing an upper limit of 8/268\,=\,3.0$\pm$1.0\% at the 1$\sigma$ level on the frequency of wide companions. We list these companion candidates in Table \ref{tab_CPM_MP:tab_wide_comp_VHS} in Appendix \ref{CPM_MP:wide_pairs_VHS_AllWISE} and plot them as blue crosses in Fig.\ \ref{fig_CPM_MP:sample}.

We repeated the search for faint companions around the individual catalogues to evaluate whether or not we could set a more stringent constraint on the multiplicity. We found 6580 stars of the 15432 covered by VHS, whose distances range from 50 to 400 pc. In this sample, we identified one bright companion detected in $Gaia$ DR3 at 4.85 arcsecs from L\,313-33 \citep{schneider16a}, whose RUWE parameter is around 1.9 , thusindicating a potential multiple system. We did not find this secondary in our original $Gaia$ sample because its metallicity is $-$0.56 dex in $Gaia$ DR3, while the primary has a metallicity of $-$1.88 dex although proper motion and parallax are in agreement for both components. The metallicity of the pair is most likely not so low because it is not in the final sample of metal-poor stars discussed in this work. At larger separations beyond 20 arcsecs from their primaries, we spotted another 12 companions also detected in $Gaia$ DR3, all with metallicities higher than the lower limit of $-$1.5 dex and outside our final sample. Three out of 12 secondaries have RUWE parameters significantly larger, pointing towards multiple systems. At fainter magnitudes ($J$\,$>$18 mag), we identified two wide companion candidates at 31.1 and 23.3 arcsecs from TYC 7650-1108-1 with [Fe/H]\,=\,$-$2.38 dex) and CD-24 17504 with [Fe/H]\,=\,$-$4.04 dex \citep{ishigaki12,yoon16}, respectively. The latter is very metal-poor primary whose companion has been identified as part of the sample of \cite{li22a} but rejected with additional GTC observations (see below). We inferred an upper limit of 0.21$\pm$0.06\% ((1+12+2-1)/6580; 1$\sigma$) on the multiplicity rate for projected separations larger 500--72000 au for the range of validity of the VHS-WISE search (10--180 arcsecs) and distance interval.


We reiterated the cross-match process for the 6500 and 1589 metal-poor stars in the catalogues of \citet{andrae23b} and \citet{zhang23a}, yielding 2671 and 665 counterparts in VHS\@. We identified 39 and 38 potential wide companions in \citet{andrae23b} and \citet{zhang23a}, respectively. After further inspection and removal of sources detected in $Gaia$ revealing distinct proper motions, we ended up with two and one companions confirmed with $Gaia$ for the samples of \citet{andrae23b} and \citet{zhang23a}. The system in common to both catalogues and our original selection is the LP\,850-23/LP\,850-24 system. The additional wide pair in the catalogue of \citet{andrae23b} is L\,158-59/L\,158-60 with a separation of about 63 arcsecs. Once again, the issue comes from the $Gaia$ metallicities of one of the components, which is above our original cut in metallicity. Hence, the success rate of truly co-moving pairs of the VHS/AllWISE search is below 2/39\,=\,5.1\% and 1/38\,=\,2.6\% for the catalogues of \citet{andrae23b} and \citet{zhang23a}, respectively.

We also identified 5 (17) and 3 (15) companion candidates without astrometric information in $Gaia$ DR3 (no $Gaia$ DR3 counterpart) in \citet{andrae23b} and \citet{zhang23a}, respectively. All are faint in the NIR, with VHS $J$-band magnitudes typically fainter than 17.5 mag, implying $G-J$ colours redder than 3.5 mag typically of ultracool dwarfs \citep{reyle18,reyle21}. Considering the low success rate identifying companions in VHS-AllWISE above, we did not expect to find many bona fide companions in those lists without $Gaia$ counterparts, a few at most. Therefore, we inferred a binary frequency of metal-poor dwarfs in the sample of \citet{andrae23b} in the range of 1/2671\,=\,0.04\% and (1+5+17)/2671\,=\,0.86\%, while those percentages are around 0.15--2.9\% for the sample from \citet{zhang23a}. We conclude that the overall binary frequency is at most 3\%, most likely below 5\% for projected separations larger than 1500--7200 au because of the success rate discussed above.
%

We repeated the same procedure to look for wide companions among the metal-poor stars in the catalogue of \citet{li22a} and the Jinabase sample covered by VHS\@. We identified a potential wide companion to 02:13:40.24 $-$00:05:18.6 in \citet{li22a}; however we rejected it because we spotted an optical counterpart in the optical and no clear motion between the Sloan digital sky survey \citep[SDSS;][]{york00} and Panoramic Survey Telescope and Rapid Response System \citep[Pan-STARRS;][]{chambers16a} detections. We did detect one possible companion at a separation of 23.7 arcsecs (23:07:22.25 $+$23.52:35.0) to CEMP-no CD-2417504 \citep{ishigaki12,yoon16} in the Jinabase sample, respectively. We obtained CCD images of this companion candidate on 21 July 2023 with Optical System for Imaging and Low Resolution Integrated Spectroscopy (OSIRIS) camera-spectrograph \citep{cepa00} in its new upgraded OSIRIS$+$ configuration\footnote{\url{https://www.gtc.iac.es/instruments/osiris+/osiris+.php}} on the Gran Telescopio de Canarias (GTC).

The OSIRIS$+$ detector is a blue sensitive monolithic 4k$\times$4k CCD, operated in 2$\times$2 binned mode giving a plate scale of 0.25"/pix and providing a unvignetted FOV of 7.8'$\times$7.8'. The standard operation mode of the instrument uses a readout speed of 233 kHz, which gives a gain of 1.9 electrons/ADU and readout noise of 4.3 electrons. We obtained a serie of 5 images using the Sloan $g$, $r$, $i$, and $z$ filters with individual exposure times of 120s with a small offset of 10 arcsecs between them, giving a total on-source time of 600s per filter. We reduced the images using an in-house pipeline for OSIRIS$+$ in imaging mode (SAUSERO), which comprises the standard procedures including bias, flat field, cosmic ray removal and a final combined image calibrated astrometrically in each passband with an average seeing of 0.9".
We did not detect any obvious motion between the OSIRIS and VHS epoch of the proposed companion to CD-2417504, yielding to the rejection of this candidate.
The statistical multiplicity results are consistent with the former search due to the much lower number statistics.

\subsection{Lucky imaging follow-up}
\label{CPM_MP:search_LI}

We carried out high spatial resolution imaging of a sub-sample of binary system
candidates from the original $Gaia$ sample of 15432 objects with FastCam \citep{oscoz08}, the "lucky imaging" instrument on the 1.52m Carlos Sanchez telescope
at Teide Observatory, Tenerife, Canary Islands in visitor mode on the nights of 18 October (observers EGE \& RC) and 14 December 2023 (observer JYZ \& RC). We note that none of these sources are common to the final sample of 610 metal-poor stars. 
We did not see any bias in the effective temperature and metallicity distribution of this sub-sample with respect to the full sample of 15432 objects. The night on 18 October was affected by clouds whose thickness and presence became more significant towards the end of the night. The seeing was variable between 1.5 and 2.5 arcsecs. The night on 14 December was dark and clear with no moon and no clouds but the peak of Geminides. The seeing was below 1.5 arcsecs all night.

The upgraded FastCam camera is equipped with a Andor iXon DU-888U3-CSO\#BV back-illuminated system\footnote{\url{http://research.iac.es/OOCC/iac-managed-telescopes/telescopio-carlos-sanchez/fastcam/}}. 
The camera has 1024$\times$1024 pixels with a size of 13 microns, low read-out noise ($<$1 electron), and more than 95\% quantum efficiency. We conducted observations in the broad-band Johnson $I$ filter with a reduced field of view of 400 pixels aside to reduce the individual on-source integration times to about 15\,ms. Typically, we selected short exposures of 15ms repeated 30000 times for each target, yielding a total on-source integration of 7.5\,min.

We targeted a total of 48 sub-dwarf candidates: 17 on 18 October and 31 on 14 December (Table \ref{tab_CPM_MP:tab_LI_targets}). We also targeted the Cyg\,22 region and the Trapezium Cluster with several objects in the field to measure accurately the pixel scale on 18 October and 14 December 2023, respectively. We measured a pixel scale of 0.03493 and 0.03498 arcsec per pixel and rotation angle of 1.033 and 1.288 degree on 18 October and 14 December 2023, respectively.

We also collected high resolution images of nine targets with FastCam on the Nordic Optical Telescope (NOT) in service mode on the night of 22 and 23 January 2024 (observer RC). Among these, four were also observed with FastCam on the Carlos Sanchez (see above), while the other five had no first epoch with FastCam. Using the Trapezium Cluster as astrometric reference, we measured a pixel scale of 0.02521 arcsec per pixel and a rotation angle of 89.48 degrees.



The data reduction of the raw images was carried out with a homemade pipeline developed specifically for FastCam. This pipeline is based on the "lucky images" technique, as described in \citet{baldwin01}, \citet{tubbs02} and references therein. Images are chosen based on their Strehl ratio, as it can be regarded as a metric for assessing the impact of atmospheric turbulences on the image. The higher the Strehl ratio, the better the quality of the image.
We selected the best 10\% images to generate the final combined image shown in Appendix \ref{CPM_MP:images_FastCam}. When the pipeline detects two stars in the image with similar magnitudes, it can utilise either of them for recentering, resulting in an image with a central star surrounded by two ghost stars. To address this issue, we devised an algorithm named "two stars", which forces the pipeline to identify the pair of brightest speckles with a fixed separation and angular orientation, using them as a reference to recenter using always the same star. In particular, we note that 0219$+$30 appears as a false triple in the TCS/FastCam images; we confirm that this system is a binary, using NOT data with improved flux and resolution. 

We measured the separation and angle of the binary systems detected in our sample with the ruler option on the {\tt{ds9}} fits viewer. The angle was calculated taking the peak of the primary as reference going from west to east. The typical uncertainty on the angle is a one degree, while the uncertainty on the separation is two tens of a pixel. We also give the separation in au by multiplying by the pixel scale of FastCam on the night of observations. We also report the ratio of the peak of the flux between the primary and the secondary in the $I$-band, which is a good proxy for the flux ratio. The flux of the primary is the taken from the pixel with the higher number of counts while the peak of the secondary comes either from the most illuminated pixel or the average of the four pixel with the largest fluxes. The typical error bar on the fluxes of the secondaries range from 20 to 200 for the best and worse cases. The targets with the largest uncertainties in all three parameters are 0927$-$08, 2007$+$13, and 2307$+$21\@. We note that 0426$+$15 might not be single while 0047$-$01 is either a binary or a triple.

%
%
\section{Discussion}
\label{CPM_MP:discussion}

The final sample is the most complete sample presented in this work and relies on $Gaia$ metallicities derived from three distinct techniques, that have indicated these sources are metal-poor with a high level of confidence. This sample contains mid-F to early-K stars with temperatures intervals of 4970-6470\,K, 5370--6660\,K, and 5380--6770\,K from \citet{andrae23b}, \citet{zhang23a}, and $Gaia$ DR3, respectively. The mean and median effective temperatures are 5773/5780\,K, 5998/6002\,K, and 6202/6271\,K, respectively, showing that there are significant offsets between the determination of \citet{andrae23b} and the other two. Naturally, the differences in temperatures impact the gravity and metallicity solutions. The lowest metallicities in \citet{andrae23b}, \citet{zhang23a}, and $Gaia$ DR3 samples are $-$2.05, $-$1.85, and $-$4.0 dex with very comparable median values of $-$2.05, $-$1.85, and $-$1.98 dex, respectively. The significant difference in the lowest metallicities is obvious in Fig.\ \ref{fig_CPM_MP:FeH_compare_main}. The gravities inferred in all catalogue are in agreement, with minimum and maximum values of $\log$(g) of 3.6 to 4.9 dex. Hence, we did not divide the sample into sub-classes.

\begin{figure*}
 \centering
 \includegraphics[width=0.32\linewidth, angle=0]{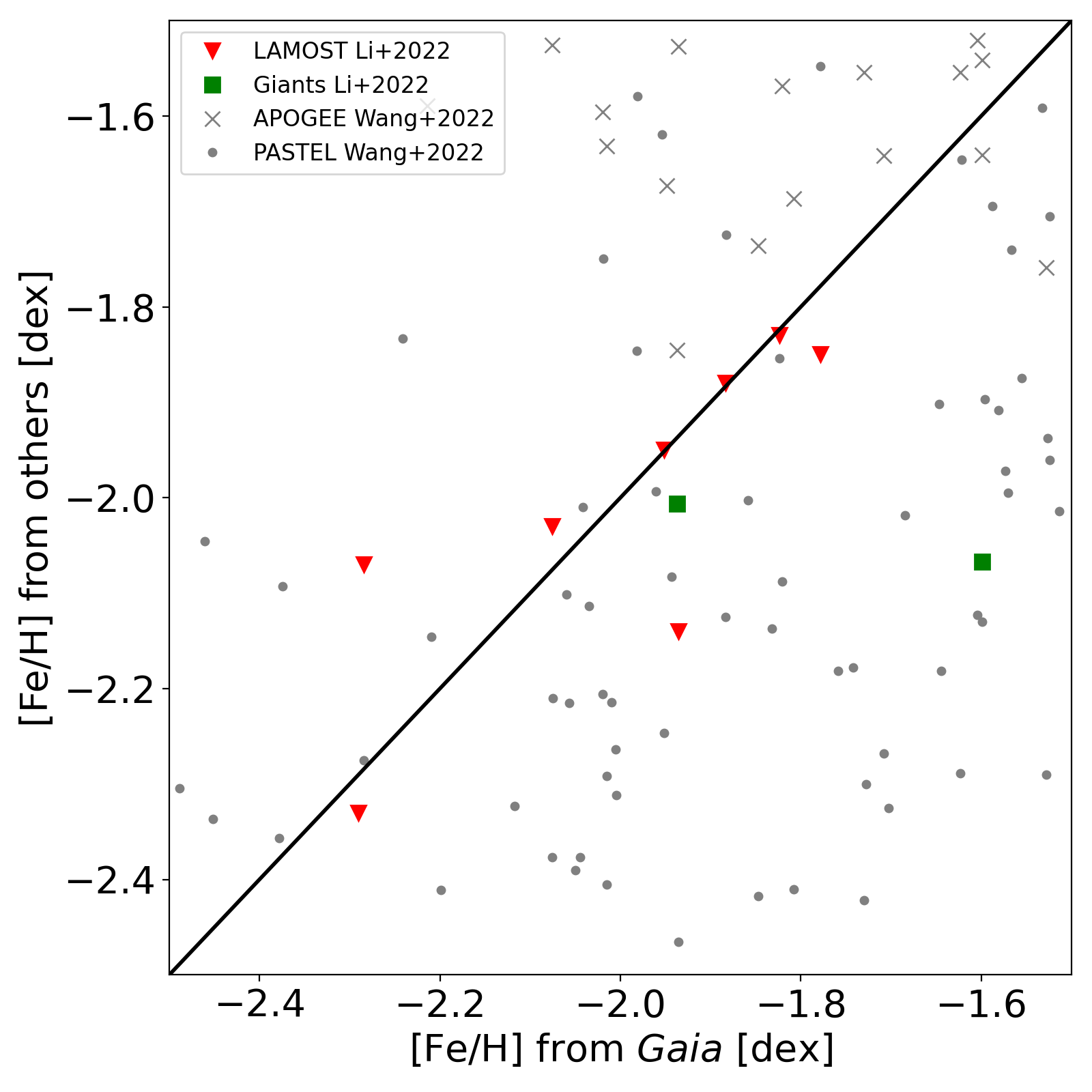}
 \includegraphics[width=0.32\linewidth, angle=0]{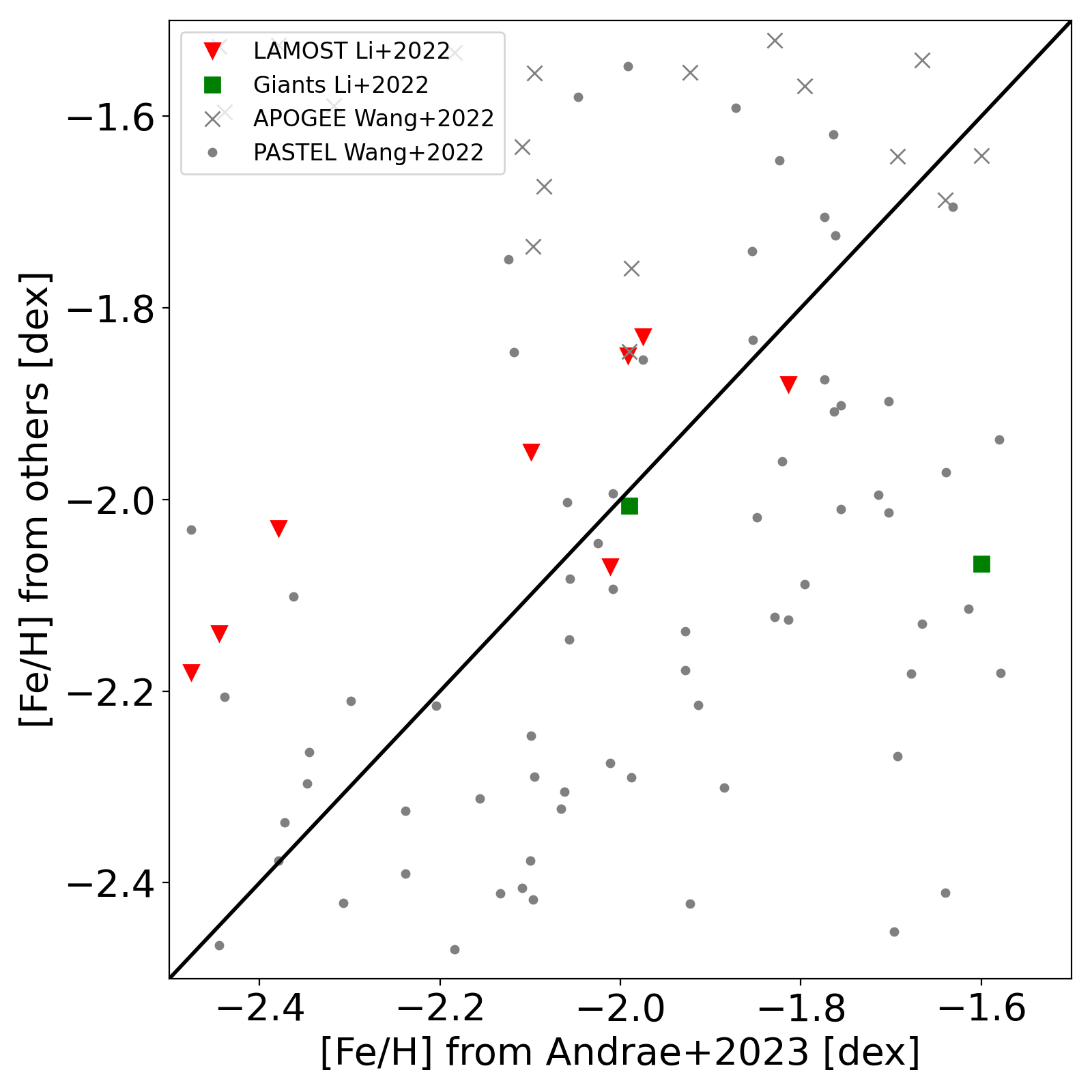}
 \includegraphics[width=0.32\linewidth, angle=0]{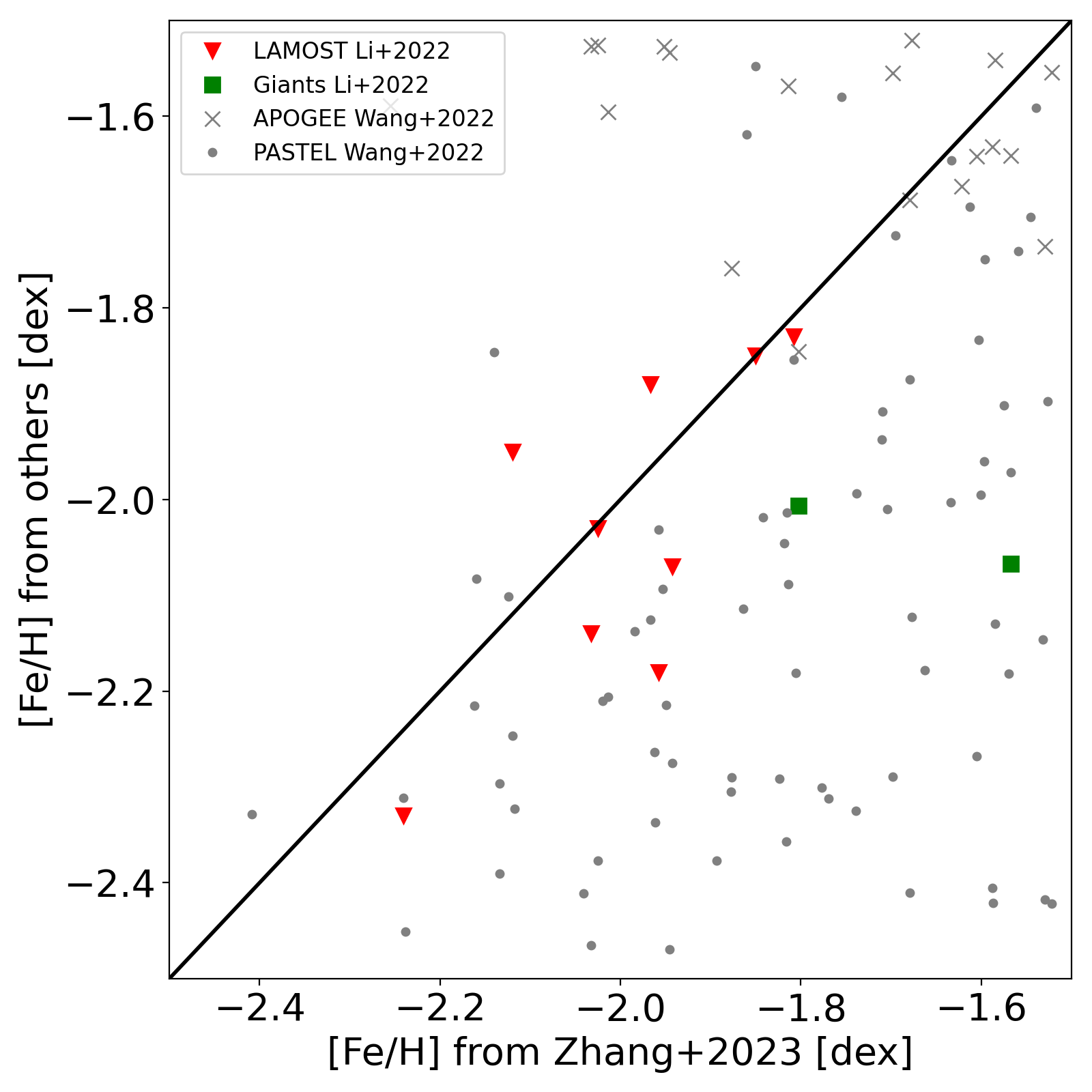}
 \caption{Comparison of metallicities from different catalogues.
 {\it{Left:}} Comparison of the $Gaia$ metallicities (x-axis) with the spectroscopic determinations of \citet{li22a}, \citet{li22b}, and \citet{wang22a}, shown as red triangles, green squares, and grey symbols, respectively.
 {\it{Middle:}} Same as middle panel but comparing the spectroscopic metallicities with the determination from \citet{andrae23b} on the x-axis.
 {\it{Right:}} Same as other panels but with metallicities from \citet{zhang23a} on the x-axis.
 }
  \label{fig_CPM_MP:FeH_compare_spec}
\end{figure*}

We cross-matched the sample of 610 sub-dwarfs with several catalogues with spectroscopically derived physical parameters, with a special focus on the metallicity (the key parameters in this work). First, we considered the 400 very metal-poor stars studied with LAMOST and Subaru \citep{li22a}, which includes 11 objects in our sample (red triangles in Fig.\ \ref{fig_CPM_MP:FeH_compare_spec}).
Second, we identified two sources in the catalogue of 1.2 million giants with abundances from LAMOST DR8 \citep{li22b}, which are plotted as green squares in Fig.\ \ref{fig_CPM_MP:FeH_compare_spec}.
Third, we added the 89 sources with metallicities from APOGEE and PASTEL in Table 1 of \citet{wang22a} and plotted as black crosses and dots in Fig.\ \ref{fig_CPM_MP:FeH_compare_spec}, respectively. We observe that the spectroscopic metallicities from APOGEE \citep{jonsson20a} tend to be lower by about 0.3 dex on average compared to the three estimates based on $Gaia$ data, while the metallicities derived using the bibliographical catalogue PASTEL \citep{soubiran10,soubiran16} appear higher \citep{wang22a}. The spectroscopic determination of \citet{li22a} agree with the photometric estimates. Overall, the spectroscopic metallicities seem to point towards lower values. We conclude that the metallicities of our sample are accurate to $\pm$0.5 dex in absolute terms, suggesting that the subsequent analysis considers truly metal-poor stars as those with metallicities below $-$1.0 dex with a high-level of confidence.

%
%
%

%
%
\section{Multiplicity of metal-poor stars}
\label{CPM_MP:multiplicity}

In this section, we discuss the multiplicity of metal-poor stars based on the samples studied in Section \ref{CPM_MP:search_Gaia}. Hence, we aim to offer a global view on the multiplicity of metal-poor solar-type stars.

At separations closer than about 20 mas, $Gaia$ is unable to separate stars. Assuming a distance interval of 50--400 pc for our final sample, this translates into projected physical separations in the  $\sim$1--8 au range. The RUWE parameter is close to 1.0 for sources well fit by a single-star model. If the RUWE is larger than 1.25 in $Gaia$ DR3 \citep{castro_ginard24}, the astrometric solution is less compatible with a single-star model, implying that the probability that a system is, in fact, a multiple would increase. Assuming that all objects with RUWE\,$\geq$1.25 are binaries or higher-order multiples in $Gaia$ DR3 (which is not necessarily the case; see below) as described in \citet{castro_ginard24}, we found 113 candidates out of 610. This would imply a multiplicity fraction of 113/610\,=\,18.5$\pm$1.7\% for separations below 1--8 au, assuming Poisson statistics. Our sample exhibits RUWE values between 1.15 and 1.36, implying an interval for the frequency of multiple systems of around 13.8--26.0\% (Fig.\ \ref{fig_CPM_MP:histogram_RUWE}). If we increase the lower limit of the RUWE parameter to 2.0, the number of sources decreases to 6.2\% (38 sources). Hence, we conclude that the frequency of multiple systems increases swiftly for separations closer than $\sim$8 au (20 mas at a maximum distance of 400 pc), yielding a minimum multiplicity rate of 20$\pm$6\%. This number might be twice as high due to the $Gaia$ incompleteness on the true fraction of binaries. If we focus on unresolved binaries ($<$\,20 mas), which are better sampled by the percent of successful image parameter determination windows {\tt{IPDfmp}} parameter \citep{lindegren21} with more than one peak (IPDfmp\,$>$\,2) of $Gaia$ \citep{tokovinin23c}, we can thereby infer a fraction of 9/610\,=\,2.0$\pm$0.5\%.

We have double-checked the multiplicity cross-matching the original sample with the catalogues of \citet{andrae23b} and \citet{zhang23a} independently. We found 1129 and 695 sources, respectively, and RUWE (IPDfmp) percentages of 18.0$\pm$1.3\% (2.2$\pm$0.4\%) and 20.7$\pm$1.7\% (3.5$\pm$0.7\%), which is consistent with the former derivation within the Poisson errors. These multiplicities agree with the frequency of solar-type stars of about 20\% for separations closer than 8 au derived from Figure 13 of \citet{raghavan10}.

At intermediate separations of a few to hundreds of AU, the fraction has been found to remain constant and independent of metallicity \citep{koehler00a,zinnecker04}. We identified nine wide co-moving pairs with separations between 1.5 and 22.5 arcsecs, corresponding to projected physical separations in the 500--8700 au considering the distance of the primaries.
We note that $Gaia$ can detect approximately 60\% of all binaries down to $G$\,=\,20 mag closer than 250 pc to the Sun. This efficiency drops to 35\% out to 1000 pc. If we interpolate linearly to the maximum distance of our sample (400 pc), the frequency would be complete to about 55\%. 
Taking into account all these uncertainties, we derive a frequency of wide companions of 9/610\,=\,1.48$\pm$0.49\% for projected separations between 8 and 10000 au (Table \ref{tab_CPM_MP:tab_binary_stats}). Applying a correction for the $Gaia$ incompleteness, this rate could be as high as 3.0\% at 3$\sigma$. This rate is significantly lower than frequency of close-in binaries among solar-type FGK dwarfs \citep[27--30\% for separations of 8--2000 au;][]{raghavan10}, denoting a sharp decrease in the multiplicity of sub-dwarfs with separation. 
%

We did find nine co-moving pairs that could be resolved on seeing-limited images because the smaller separation is 1.5 arcsecs (corresponding to about 500 au). This is at odds with the seeing-limited deep imaging surveys of metal-poor stars \citep{allen00,zapatero04a} suggesting a fraction of $\sim$15\% and solar-type stars \citep[10.8\% for a$>$1000 au;][]{raghavan10}. One possible explanation for this is that we have specifically focussed on stars with metallicities below $-$1.5 dex. \citet{allen00} identified 122 wide binaries among about 1200 metal-poor stars \citep{schuster88,schuster89,schuster93a} but only seven binaries have metallicities below $-$1.5 dex (see Table 2 of their appendix). Among these, the separations of the three most metal-poor stars with [Fe/H] below $-$2.0 dex lie between 32000 and 125000 au while the remaining ones are closer, with separations between 530 and 8600 au. \citet{zapatero04a} searched for companions up to 25 arcsecs from a sample 473 metal-poor stars, with about half of them displaying metallicities below $-$1.5 dex. These authors reported six previously known companions and five new companions with separations between 37 and $\sim$56500 au, yielding a frequency of about 4.5\%. Many stars with [Fe/H] below $-$2.0 dex also show separations larger than 20000 au, as in the sample of \citet{allen00}. The wide binary frequency studied for a wide range of metallicities from $-1.5$ dex to $+$0.5 dex at separations of 10000--100000 au combining LAMOST spectra with astrometry from $Gaia$ DR2 concluded that there must be a decrease in the multiplicity fraction with metallicity, reaching about 1\% at [Fe/H]\,=\,$-$1.5 dex \citep{hwang21a}, which is consistent with our results. Our sample is the largest homogeneous sample of metal-poor stars with metallicities below $-$1.5 dex with a dedicated analysis of the multiplicity. 
One possible explanation for the difference with earlier studies may come from the $Gaia$ metallicities that may differ between the primary and secondary, as in the case of three of the nine pairs listed in Table \ref{tab_CPM_MP:tab_wide_comp} in Appendix \ref{CPM_MP:wide_pairs_300arcsec}, combined with the limited sensitivity of $Gaia$ to very cool companions.

Overall, we find a sharp decrease in the frequency of multiple systems beyond 8 au (assuming the maximum distance of 400 pc for our sample). On the one hand, this decrease appears sharper than for solar-type stars. On the other hand, the frequency is significantly lower for metal-poor solar-type stars than their solar-metallicity counterparts (3.0\% at 3$\sigma$ confidence vs 10--30\% depending on the separation).
%
However, the FastCam follow-up of 48 sub-dwarfs shows that a high value of the $Gaia$ RUWE parameter does not always correlate with multiplicity. We find that 62.5$\pm$2\% of 48 sub-dwarfs are multiples in a range of projected separations of 105--840 au. Looking at the full sample of targets imaged with FastCam, we find that the numbers of resolved systems among low and high RUWE is comparable (48\% vs 52\%), demonstrating that dedicated high-resolution imaging campaigns are needed to shed more light on the multiplicity as a function of metallicity and projected separations. These numbers suggest that the field population might rather come from low density environments rather than dense regions, as seen in open clusters and star-forming regions \citep*{deacon20}.

%
%
\begin{figure*}
   \centering
   \includegraphics[width=0.32\linewidth, angle=0]{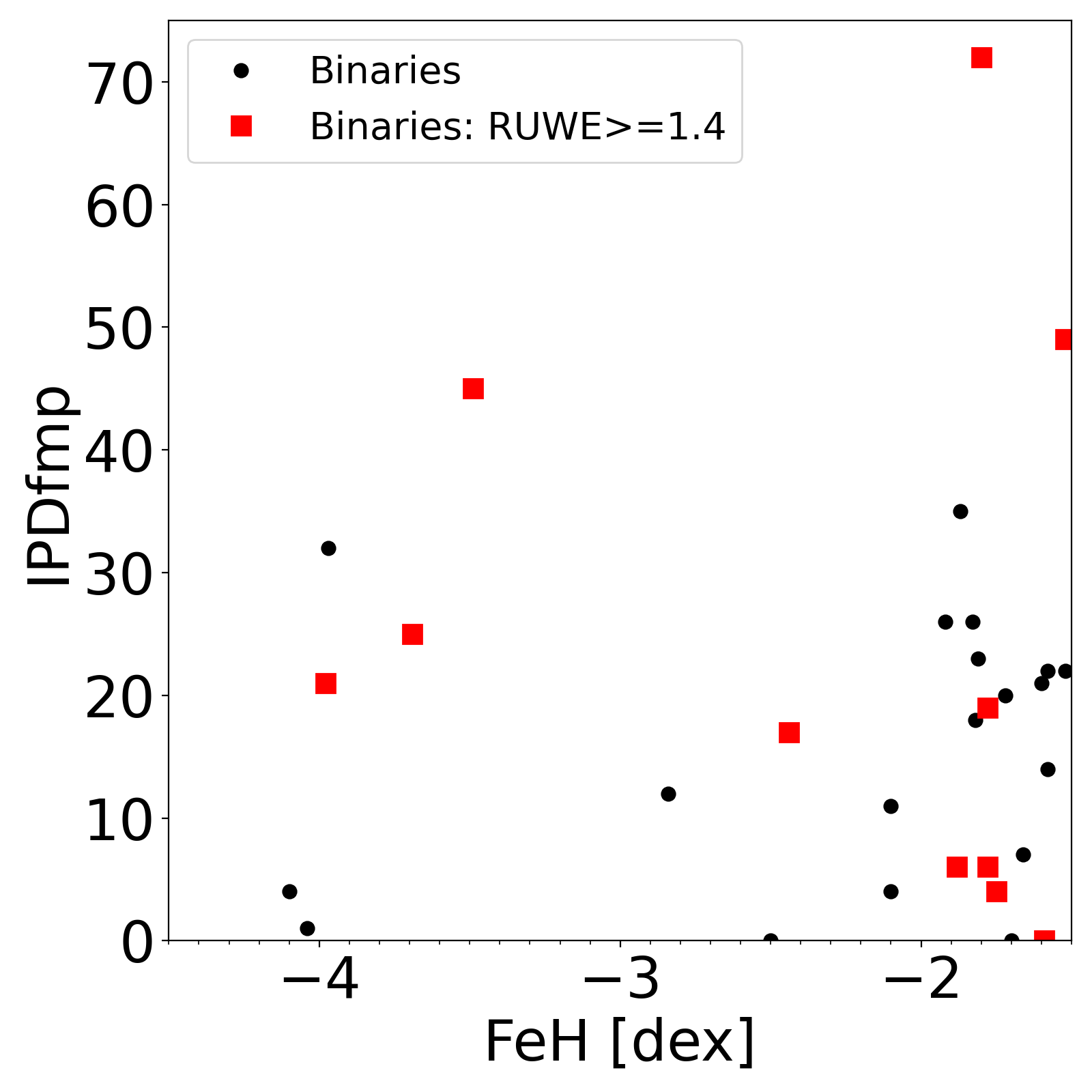}
   \includegraphics[width=0.32\linewidth, angle=0]{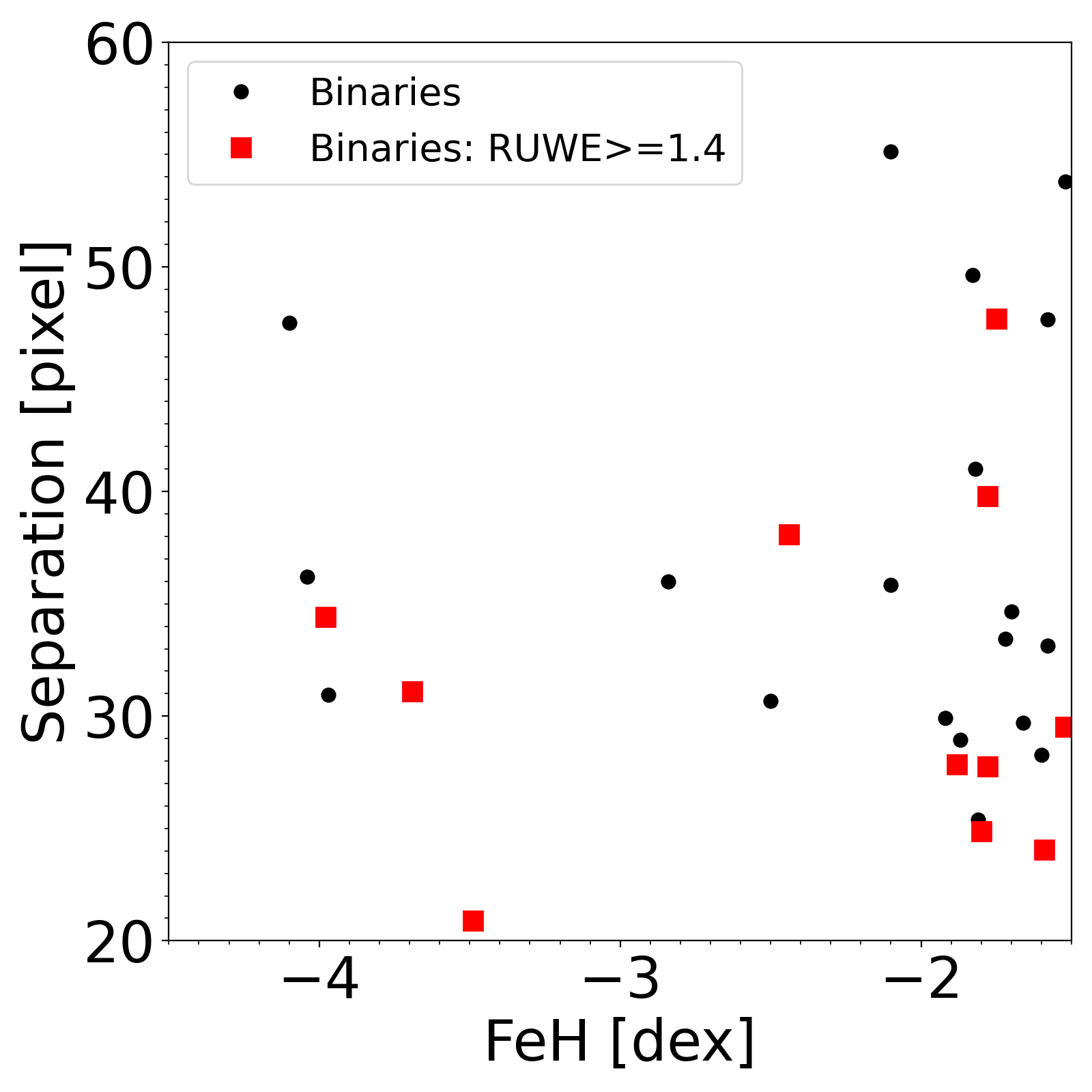}
   \includegraphics[width=0.32\linewidth, angle=0]{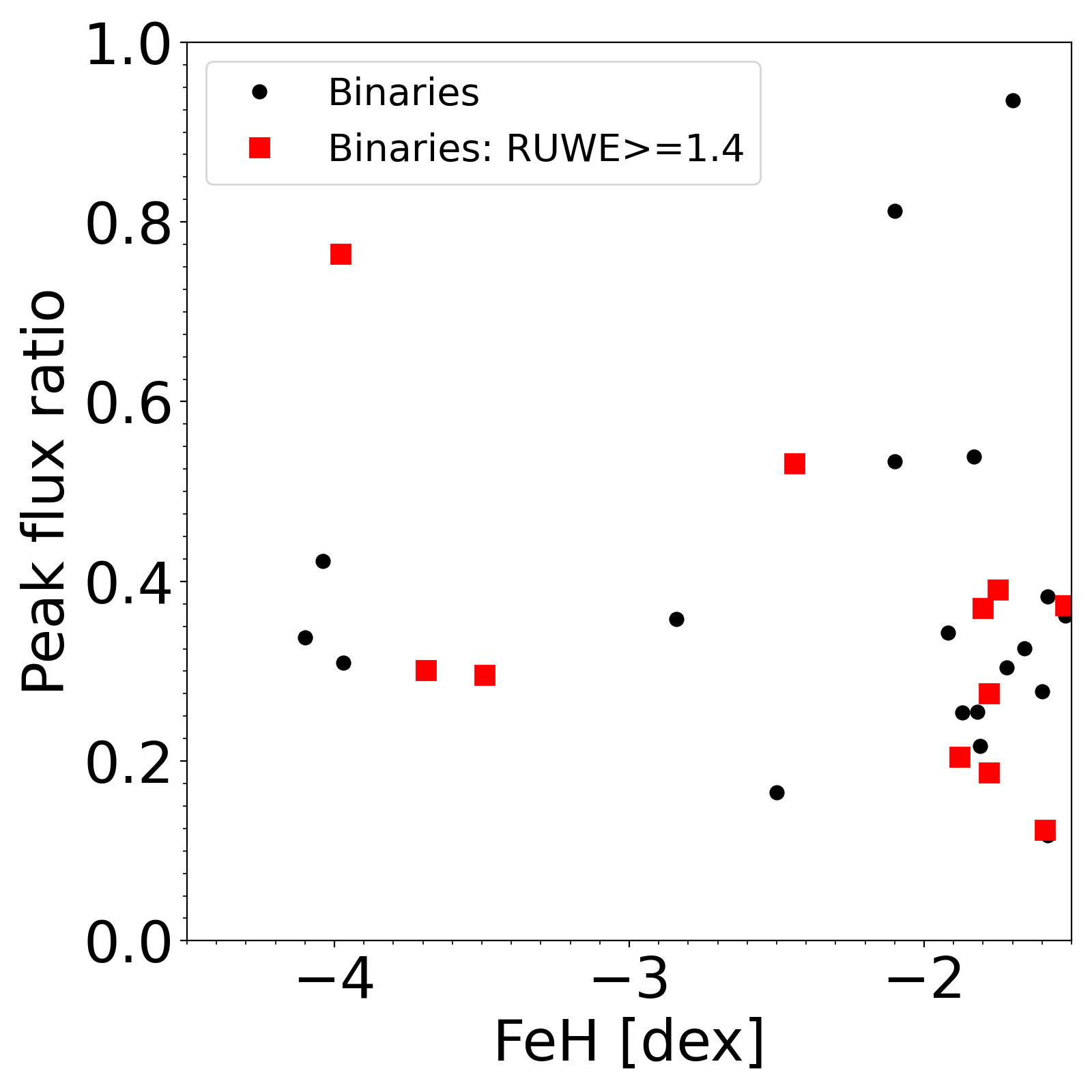}
   \includegraphics[width=0.32\linewidth, angle=0]{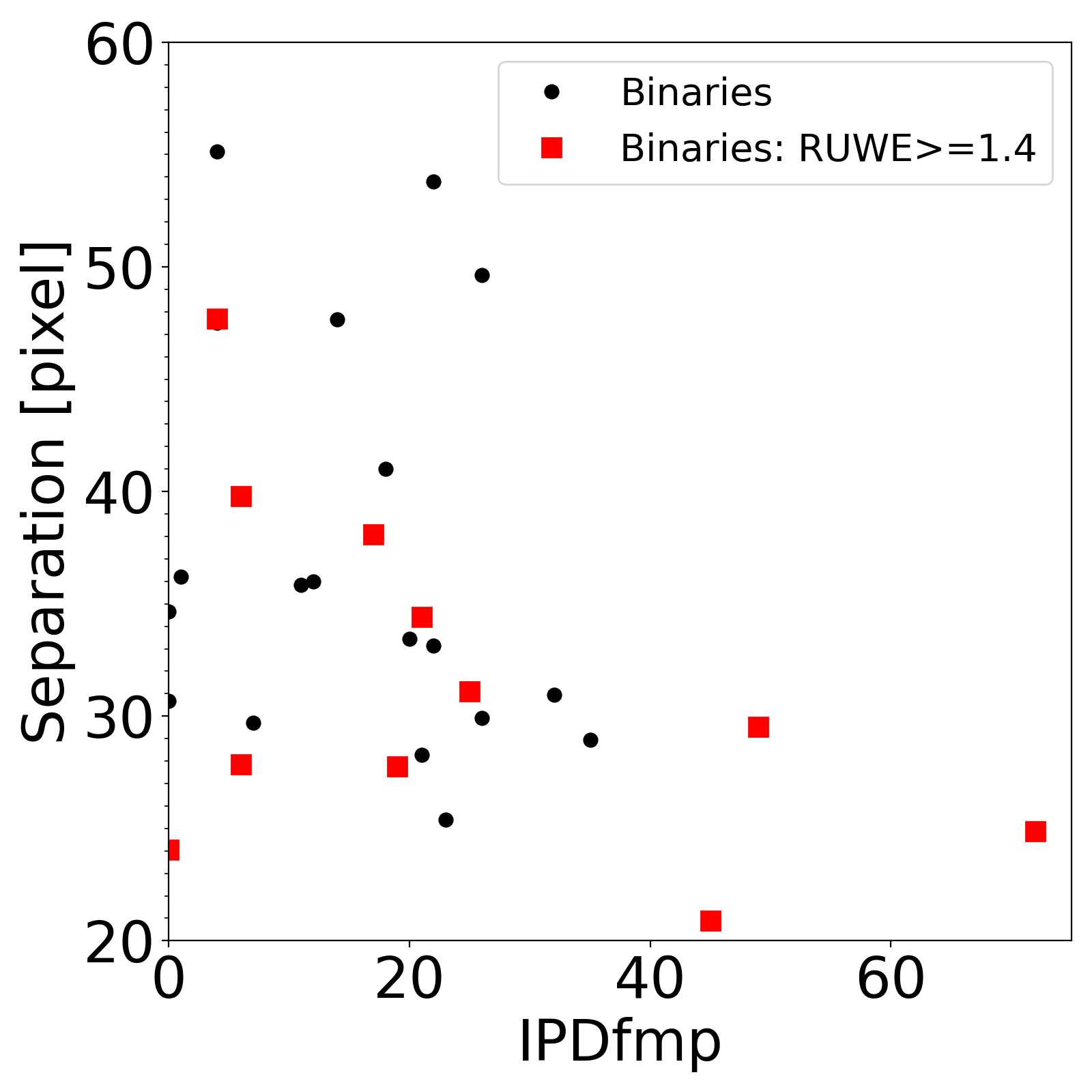}
   \includegraphics[width=0.32\linewidth, angle=0]{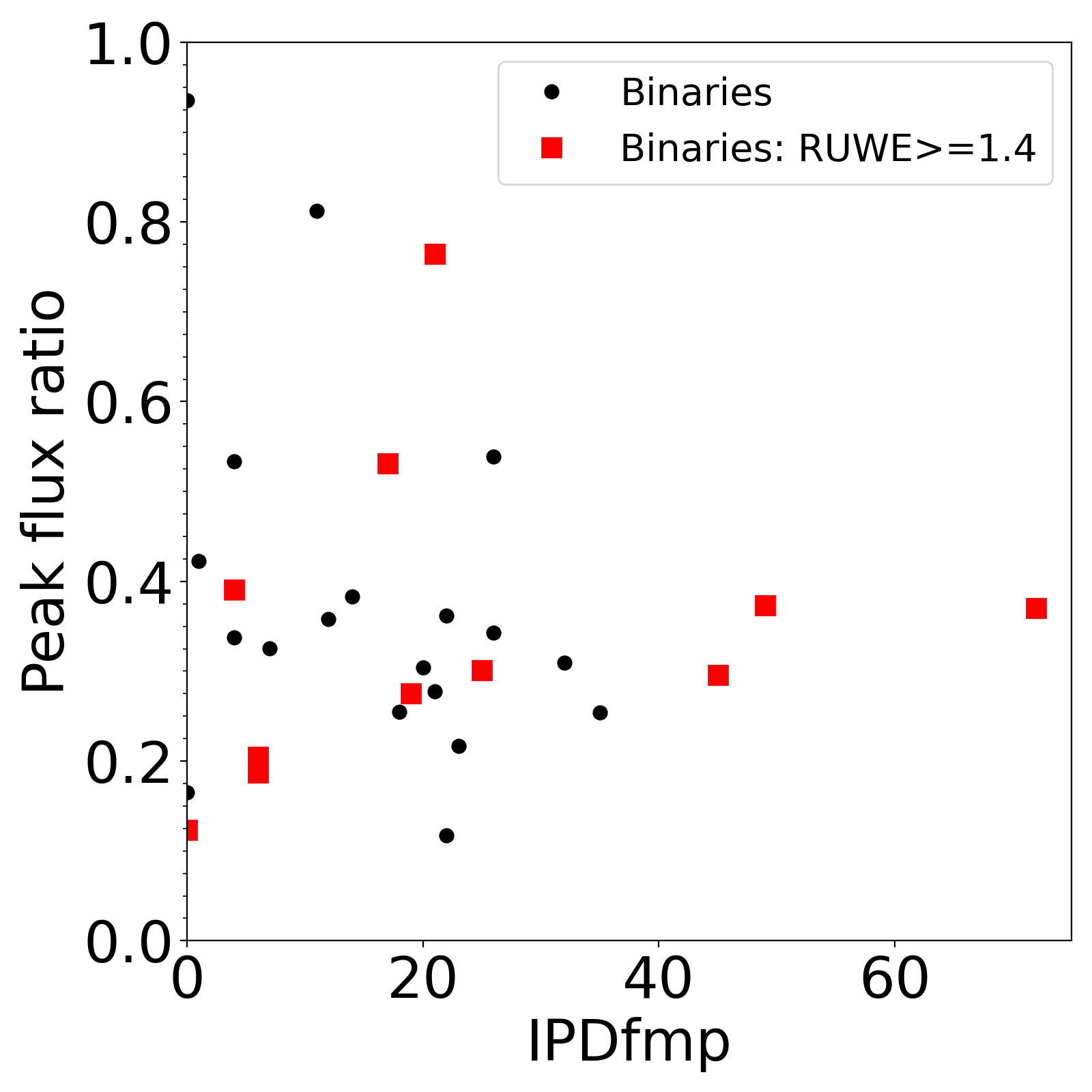}
   \caption{Properties of the sub-dwarf systems resolved with FastCam on the 1.5-m Carlos S\'anchez telescope.
   }
   \label{fig_CPM_MP:plots_FastCam}
\end{figure*}
%

%
%
\begin{table}
 \centering
 \caption[]{Multiplicity statistics of our samples.
 }
 \small
 \begin{tabular}{@{\hspace{0mm}}c @{\hspace{2mm}}c @{\hspace{2mm}}c @{\hspace{2mm}}c @{\hspace{2mm}}c @{\hspace{2mm}}c @{\hspace{2mm}}c@{\hspace{0mm}}}
 \hline
 \hline
Sep           &      Sep         &  SpT      & N$_{tot}$ &  N$_{comp}$ &   Freq   & Catalogues \cr
 \hline
 arcsec       &       au         &           &           &             &    \%    &            \cr
 \hline
$\leq$\,0.02  &  $\leq$\,1--8      &   Any   &  610      &     113     &    18.7$\pm$1.7  & $Gaia$  \cr
$\leq$\,0.02  &  $\leq$\,1--8      &   Any   &  1129     &     204     &    18.3$\pm$1.3  & Andrae$+$2023  \cr
$\leq$\,0.02  &  $\leq$\,1--8      &   Any   &   695     &     144     &    20.7$\pm$1.7  & Zhang$+$2023  \cr
25--300     &    10000--120000      &   Any   &  610      &       0     &    $<$0.6\%  & $Gaia$  \cr
0.02--25      &     8--10000       &   Any   &  610      &       9     &  1.5$\pm$0.5\%  & $Gaia$  \cr
10--180       &   1500--72000      &   Any   &   268     &       8     &  3.0$\pm$1.0\%  & VHS-AllWISE  \cr
\hline
0.7--2.1     &    100--840        &   Any   &    48     &      30     &   62.5  & FastCam  \cr
10--180      &   500--72000       &   Any   &   143     &       0     &  $<$0.7  & Jinabase  \cr
10--180      &   800--28800       &   Any   &    10     &       0     &  $<$10  & Li$+$2022  \cr
\hline
 \label{tab_CPM_MP:tab_binary_stats}
 \end{tabular}
 \tablefoot{The top part shows samples with a 1$\sigma$ confidence level: separations, spectral types, numbers of sources and companions, frequency. The bottom part corresponds to small or biased samples with least statistics.}
\end{table}

Among the sample of 49 sub-dwarfs targeted with FastCam, we resolved 30 systems with separations ranging between 0.7 and 2.1 arcsecs, corresponding to projected separations between 21 and 840 au considering the range of distances of this FastCam sample (30--400 pc). We note that 13 of the 30 resolved systems had a RUWE higher than 1.4 (43.3\%), while only 5 of the 18 unresolved systems had RUWE below 1.4 (27.8\%). We emphasise that this lower limit for the RUWE parameter applied for $Gaia$ DR2 prior to the updated analysis of $Gaia$ DR3 by \citet{castro_ginard24}. The derived frequency is high with 63.2\% of multiple systems with an uncertainty of about 2\% because the multiplicity of one target (2307$+$21) is dubious. We emphasise that this FastCam sample is highly biased because we focussed mainly on candidates with a high probability of being multiples. Therefore, the derived frequency must be interpreted as a strict upper limit in an incomplete sample. 

We plot several parameters in Fig.\ \ref{fig_CPM_MP:plots_FastCam}, including the $Gaia$ RUWE and IPDfmp versus the physical separation in pixels and the ratio of the peak fluxes of the secondary divided by the primary. We do not observe any significant trend in the physical separation or flux ratios as a function of metallicity. The majority of systems seem to be concentrated around separation of 30--40 pixels (1.05--1.40 arcsecs) and flux ratios of 0.2--0.4\@. However, we do observe a decline in the measured separation with increasing IDPfmp (lower right panel in Fig.\ \ref{fig_CPM_MP:plots_FastCam}). We cross-matched the list of systems observed with FastCam with $Gaia$ DR3 to figure out how many pairs our original sample based on a strict limit on the metallicity may have lost. We find that 19 had no companion reported in $Gaia$ DR3, 16 have a companion without effective temperature and metallicity estimates, 7 have a nearby object without parallax nor proper motion, 1 has two companions, 6 have more than one companion, whose metallicities are greater than $-$1.5 dex, but lower than $-$0.5 dex; this indicates that they are metal-poor, but with a metallicity discrepant between the primary and secondary. Among the pairs resolved with FastCam, most of them were indeed possible binaries in $Gaia$ DR3, except for a handful of exceptions marked in Table \ref{tab_CPM_MP:tab_LI_targets}. We conclude that the multiplicity derived from the $Gaia$ sample for separations below 25 arcsecs is a lower limit because we are missing pairs due to various incompleteness factors; in particular, discrepant metallicities between both components of the system as well as secondaries with no parallax and proper motion measurements. 

The multiplicity statistics set from the catalogues of \citet{li22a} and Jinabase sample agree with those extracted from the $Gaia$ only sample, but they have a lower level of significance due to the smaller numbers of sources in the input catalogues (10 and 143, respectively). We inferred upper limits on the former of 10$\pm$5\% and of 0.70$\pm$0.06\% for projected separations of 50--4000 au and 80--1600 au, respectively. 

We compare the multiplicity of the most statistically-significant sample with previous studies of metal-poor stars published in the literature.
At close separations ($<$8 au) based on the RUWE, our fractions of about 20\% agree with the results of long-term radial velocity surveys of large samples of truly metal-poor stars \citep[15\%;][]{partridge67,stryker85,carney94,latham02}. We conclude that initial conditions in the molecular clouds do not seem to evolve significantly over time. 

\section{Conclusions}
\label{CPM_MP:conclusions}

We present dedicated searches for wide companions to three samples of stars with metallicities below [Fe/H]\,=\,$-$1.5 dex to look for low-mass companions and infer the multiplicity statistics. The main results of our analysis are summarised below:

\begin{itemize}
\item [$\bullet$] Our sample involves 610 stars with metallicities below $-$1.5 dex and distances between 70 and 400 pc in common to three catalogues with independent determinations of the metallicities.
\item [$\bullet$] For our sample of sub-dwarfs, we infer a minimum binary fraction of 20$\pm$6\% based on the RUWE parameter for projected separations closer than $\leq$8 au, assuming a maximum distance of 400 pc.
\item [$\bullet$] Nine wide pairs separated by 1.5--25 arcsecs are detected in the $Gaia$ DR3 sample (500--8700 au), implying a fraction below 1.5$\pm$0.5\% with an upper limit of 3\% at 3$\sigma$ confidence level at odds with former studies. However, this fraction is a very likely lower limit because of various incompleteness factors that could be up to a factor of two due the completeness and the determination of metallicities from $Gaia$.
%
%
\item [$\bullet$] No pair with $Gaia$ astrometry was found with separations of 25--300 arcsecs (projected separations of 3750--120000 au), yielding a very low fraction below 1\% much lower with previous seeing-limited searches (4--5\%).
\item [$\bullet$] Several wide common proper motion companion candidates have been identified in the 10--180 arcsecs range in the VHS-AllWISE cross-match for three different samples of metal-poor stars with precise astrometry from $Gaia$. However, they require an additional astrometric and spectroscopic confirmation. We infer a multiplicity frequency below 3.0$\pm$1.0\% (1$\sigma$ confidence).
\item [$\bullet$] A sub-sample of systems with indication of multiple pairs targeted with lucky imaging revealed that 62.5\% of them are pairs in the 105--840 au separation range, with statistics relatively independent of the RUWE parameter. There is a visible trend whereby the physical separation decreases with an increasing $Gaia$ parameter {\tt{IPDfmp}}. 
\item [$\bullet$] High-resolution surveys are needed to complement the $Gaia$ data and further constrain the impact of key astrometric parameters that are most sensitive to binarity.
\end{itemize}

The main conclusions of our study reveals that short-period binaries do not seem to be impacted by metallicity or by any of the specific conditions present in the original molecular clouds. {The fraction of binaries with separations beyond 8 au appear less frequent at metallicities below $-$1.5 dex than at solar metallicity, pointing towards a significant disruption of the systems with less binding energy that is due to their old(er) age and increased probability of interactions. Complementary higher resolution imaging studies are needed to further constrain the multiplicity of this metal-poor population.

%
%
\section{Data availability}
Images of the targets observed with the FastCam instrument are available on zenodo at the URL \url{https://zenodo.org/records/14558611}.

%
%
\begin{acknowledgements}
NL and JYZ acknowledge support from the Agencia  Estatal de Investigaci\'on del Ministerio de Ciencia e Innovaci\'on (AEI-MCINN) under grants PID2019-109522GB-C53 and PID2022-137241NB-C41\@. 
APG acknowledges support  from the grant PID2020-120052GB-I00 financed by MCIN/AEI/10.13039/501100011033\@.
ELM is supported by the European Research Council Advanced grant SUBSTELLAR, project number 101054354\@.
We would like to thank the referee, Dr.\ Haijun Tian, for his comments on the original manuscript that led to a stronger and compelling analysis.
We thank Alfred Castro-Ginard for sharing his latest GUMS simulation with us to look at the distribution across the sky \citep{castro_ginard24}.
This article is based on observations made with the 1.52m Carlos S\'anchez telescope operated on the island of Tenerife by the IAC in the Spanish Observatorio del Teide (Canary Islands).
This research has made use of the Simbad and Vizier databases, and the Aladin sky atlas operated
at the centre de Donn\'ees Astronomiques de Strasbourg (CDS), and
of NASA's Astrophysics Data System Bibliographic Services (ADS).
This work presents results from the European Space Agency (ESA) space mission Gaia. Gaia data are being processed by the Gaia Data Processing and Analysis Consortium (DPAC). Funding for the DPAC is provided by national institutions, in particular the institutions participating in the Gaia MultiLateral Agreement (MLA). The Gaia mission website is \url{https://www.cosmos.esa.int/gaia}. The Gaia archive website is \url{https://archives.esac.esa.int/gaia}.
Based on observations obtained as part of the VISTA Hemisphere Survey, ESO Progam, 179.A-2010 (PI: McMahon)
Based on observations made with the Gran Telescopio Canarias (GTC), installed at the Spanish Observatorio del Roque de los Muchachos of the Instituto de Astrof\'isica de Canarias, on the island of La Palma. This work is (partly) based on data obtained with the instrument OSIRIS$+$, built by a Consortium led by the Instituto de Astrof\'isica de Canarias in collaboration with the Instituto de Astronom\'ia of the Universidad Aut\'onoma de M\'exico. OSIRIS was funded by GRANTECAN and the National Plan of Astronomy and Astrophysics of the Spanish Government.
Based on observations made with the Nordic Optical Telescope, owned in collaboration by the University of Turku and Aarhus University, and operated jointly by Aarhus University, the University of Turku and the University of Oslo, representing Denmark, Finland and Norway, the University of Iceland and Stockholm University at the Observatorio del Roque de los Muchachos, La Palma, Spain, of the Instituto de Astrofisica de Canarias.
This publication makes use of data products from the Wide-field Infrared Survey Explorer, which
is a joint project of the University of California, Los Angeles, and the Jet Propulsion
Laboratory/California Institute of Technology, and NEOWISE, which is a project of the Jet
Propulsion Laboratory/California Institute of Technology. WISE and NEOWISE are funded
by the National Aeronautics and Space Administration. \\
Guoshoujing Telescope (the Large Sky Area Multi-Object Fiber Spectroscopic Telescope LAMOST) is a National Major Scientific Project built by the Chinese Academy of Sciences. Funding for the project has been provided by the National Development and Reform Commission. LAMOST is operated and managed by the National Astronomical Observatories, Chinese Academy of Sciences. 
\end{acknowledgements}

%
%

%
%
\bibliographystyle{aa} 
\bibliography{biblio_old} 

%
%
\begin{appendix} 
\onecolumn

\section{ADQL query in $Gaia$ archive}
\label{CPM_MP:ADQLquery}

SELECT ra, dec, l, b, parallax, parallax\_error, pm, pmra, pmdec, ruwe, astrometric\_excess\_noise, astrometric\_excess\_noise\_sig, phot\_g\_mean\_mag, phot\_bp\_mean\_mag, phot\_rp\_mean\_mag, radial\_velocity, teff\_gspphot, logg\_gspphot, mh\_gspphot \newline
FROM gaiadr3.gaia\_source \newline
WHERE parallax $\geq$ 2.5 \newline
AND parallax\_over\_error $\geq$ 3. \newline
AND sqrt(pmra*pmra+pmdec*pmdec)*4.74/parallax >= 60.0 \newline
AND mh\_gspphot BETWEEN $-$10.0 AND $-$1.5 \newline
ORDER BY ra,dec \newline

\section{List of wide pairs up to 300 arcsecs with some Gaia parameters}
\label{CPM_MP:wide_pairs_300arcsec}
%

%
%
%
%
\begin{table*}[!h]
 \centering
 \caption[]{List of wide pairs up to 300 arcsecs with some $Gaia$ parameters. 
 }
\small
 \begin{tabular}{@{\hspace{0mm}}c @{\hspace{2mm}}c @{\hspace{2mm}}c @{\hspace{2mm}}c @{\hspace{2mm}}c @{\hspace{2mm}}c @{\hspace{2mm}}c @{\hspace{2mm}}c @{\hspace{2mm}}c @{\hspace{2mm}}c @{\hspace{2mm}}c @{\hspace{2mm}}c @{\hspace{2mm}}c @{\hspace{2mm}}c@{\hspace{0mm}}}
 \hline
 \hline
R.A. &  dec   &  $G$   &  Plx & $\mu_{\alpha}\cos\delta$ & $\mu_{\delta}$ &  RV  &  T$_{\rm eff}$ & [Fe/H]  & $\log(g)$ &  (A$_{G}$) & RUWE & IPDfmp & Sep \cr
 \hline
     &        &  mag   &  mas & mas/yr & mas/yr &  km/s &   K  &  dex  & dex &  mag &   &  & arcsec \cr
 \hline
00:20:04.14 & +42:43:43.7 & 12.919 & 2.600 & 183.389 & $-$172.499 & $-$191.5 & 6331 & $-$2.69 & 4.5 & 0.18 & 1.065 &  0 & --- \cr 
00:49:20.33 & +06:25:09.7 & 12.729 & 2.653 & 115.959 & $-$74.819 & $-$183.7 & 6286 & $-$2.05 & 4.4 & 0.14 & 0.999 &  0 & --- \cr 
07:18:18.28 & $-$21:56:40.3 & 13.283 & 3.042 & 45.382 & $-$54.195 & 174.9 & 5755 & $-$1.51 & 4.7 & 0.12 & 0.997 &  0 & --- \cr 
14:06:56.00 & +80:57:10.7 & 11.485 & 3.404 & $-$53.029 & $-$125.408 & $-$306.6 & 6460 & $-$2.17 & 4.2 & 0.08 & 0.971 &  0 & --- \cr 
14:53:10.23 & +48:14:43.8 & 10.761 & 3.992 & $-$13.819 & $-$273.693 & $-$156.7 & 6401 & $-$1.81 & 4.1 & 0.11 & 0.838 &  0 & --- \cr 
16:00:09.10 & +57:18:02.8 & 11.845 & 2.898 & 62.418 & $-$96.468 & $-$120.1 & 6536 & $-$2.34 & 4.3 & 0.09 & 0.913 &  1 & --- \cr 
18:21:46.55 & $-$11:40:11.7 & 12.497 & 2.891 & $-$44.518 & $-$155.644 & $-$105.6 & 6476 & $-$3.23 & 4.2 & 0.78 & 0.902 &  0 & --- \cr 
21:12:34.36 & +35:33:06.7 & 12.703 & 2.539 & $-$29.953 & $-$54.124 & $-$304.4 & 6217 & $-$1.56 & 4.4 & 0.13 & 0.939 &  1 & --- \cr 
22:03:13.73 & $-$01:13:15.2 & 11.223 & 5.803 & 197.956 & $-$127.597 & $-$52.2 & 5866 & $-$1.60 & 4.5 & 0.01 & 0.878 &  0 & --- \cr 
\hline
00:20:04.14 & +42:43:43.7 & 17.780 & 2.470 & 183.514 & $-$172.539 & 0.0 &    0 & 0.00 & 0.0 & 0.00 & 1.035 &  1 & 22.549 \cr 
00:49:20.33 & +06:25:09.7 & 11.907 & 2.573 & 115.869 & $-$74.757 & $-$188.2 & 7208 & $-$2.37 & 4.5 & 0.14 & 1.740 &  0 & 20.655 \cr 
07:18:18.28 & $-$21:56:40.3 & 18.710 & 3.092 & 45.183 & $-$54.245 & 0.0 &    0 & 0.00 & 0.0 & 0.00 & 1.244 &  0 & 1.529 \cr 
14:06:56.00 & +80:57:10.7 & 17.798 & 3.584 & $-$52.710 & $-$125.792 & 0.0 &    0 & 0.00 & 0.0 & 0.00 & 1.076 &  4 & 4.281 \cr 
14:53:10.23 & +48:14:43.8 & 18.663 & 4.024 & $-$14.584 & $-$274.064 & 0.0 &    0 & 0.00 & 0.0 & 0.00 & 1.076 &  1 & 4.931 \cr 
16:00:09.10 & +57:18:02.8 & 16.839 & 2.923 & 63.003 & $-$96.455 & 0.0 & 4036 & $-$1.30 & 4.9 & 0.36 & 2.040 &  0 & 5.640 \cr 
18:21:46.55 & $-$11:40:11.7 & 17.238 & 2.777 & $-$44.594 & $-$155.536 & 0.0 &    0 & 0.00 & 0.0 & 0.00 & 0.971 &  0 & 2.411 \cr 
21:12:34.36 & +35:33:06.7 & 18.129 & 2.454 & $-$30.142 & $-$53.516 & 0.0 & 3662 & $-$0.61 & 4.9 & 0.11 & 1.203 &  0 & 4.556 \cr 
22:03:13.73 & $-$01:13:15.2 & 16.357 & 5.794 & 197.900 & $-$127.786 & 0.0 & 3699 & $-$1.28 & 4.9 & 0.29 & 0.895 &  0 & 5.680 \cr 
\hline
 \label{tab_CPM_MP:tab_wide_comp}
 \end{tabular}
 \tablefoot{Notes: The primaries (top) and secondaries (bottom) of each pair are plotted as black crosses and black stars in Fig.\ \ref{fig_CPM_MP:sample}. The physical parameters of the primary are taken from \citet{andrae23b} while those of the secondary come from $Gaia$ DR3\@. }
\end{table*}
\section{List of wide pairs in VHS and AllWISE. }
\label{CPM_MP:wide_pairs_VHS_AllWISE}
\begin{table*}[!h]
 \centering
 \caption[]{List of wide pairs up to 180 arcsecs identified in VHS and AllWISE\@.
 }
\footnotesize
 \begin{tabular}{@{\hspace{0mm}}c @{\hspace{1mm}}c @{\hspace{2mm}}c @{\hspace{2mm}}c @{\hspace{2mm}}c @{\hspace{2mm}}c @{\hspace{2mm}}c @{\hspace{2mm}}c @{\hspace{2mm}}c @{\hspace{2mm}}c @{\hspace{2mm}}c @{\hspace{1mm}}c @{\hspace{2mm}}c @{\hspace{2mm}}c @{\hspace{2mm}}c@{\hspace{0mm}}}
 \hline
 \hline
R.A. &  dec   &  $G$   &  Plx & $\mu_{\alpha}\cos\delta$ & $\mu_{\delta}$ &  RV  &  T$_{\rm eff}$ & $\log(g)$ & [Fe/H] &  RA$_{\rm VHS}$ & dec$_{\rm VHS}$  & pmRA & pmDEC & Sep \cr
 \hline
hh:mm:ss.ss &  $^{\circ}$:':''  &   mag   &  mas   &  mas/yr   & mas/yr  &  km/s  &  K   &  dex   &   dex  &  hh:mm:ss.ss & $^{\circ}$:':'' & mas/yr & mas/yr & arcsec \cr
 \hline
03:09:36.92 & $-$43:44:47.3 & 10.987 & 3.98 & 236.74 & $-$51.87 & 430.77 & 6447 & 4.18 & $-$1.95 & 03:09:30.20 & $-$43:44:10.4 & 203.13 & $-$48.10 & 81.60 \cr 
03:18:05.91 & $-$17:38:14.7 & 10.133 & 6.69 & 28.02 & $-$84.03 & 54.46 & 6442 & 4.27 & $-$2.36 & 03:18:00.25 & $-$17:39:30.9 &  29.93 & $-$95.09 & 111.07 \cr 
05:38:20.79 & $-$44:10:36.9 & 10.622 & 8.50 & 114.08 & 133.21 & 170.88 & 6018 & 4.54 & $-$1.61 & 05:38:31.26 & $-$44:12:44.4 & 122.20 & 136.49 & 170.11 \cr 
06:33:28.82 & $-$32:45:36.1 & 13.140 & 3.50 & $-$16.29 & 131.83 & 237.10 & 5799 & 4.69 & $-$2.16 & 06:33:23.16 & $-$32:44:40.3 & $-$16.13 & 133.31 & 90.62 \cr 
14:12:01.81 & $-$18:48:42.6 & 11.900 & 4.46 & $-$220.42 & $-$101.38 & 33.26 & 6047 & 4.48 & $-$1.53 & 14:11:59.07 & $-$18:50:16.3 & $-$211.90 & $-$97.99 & 101.49 \cr 
20:47:12.42 & $-$10:11:36.1 & 11.914 & 4.64 & 28.28 & $-$316.56 & $-$113.31 & 5667 & 4.46 & $-$1.83 & 20:47:01.85 & $-$10:11:58.3 &  28.03 & $-$344.55 & 157.61 \cr 
21:39:03.40 & $-$71:21:34.3 & 11.485 & 2.96 & 99.39 & $-$64.78 & 47.10 & 6434 & 4.18 & $-$1.89 & 21:39:23.88 & $-$71:23:34.8 &  99.25 & $-$58.47 & 155.39 \cr 
22:55:18.93 & $-$40:36:42.2 & 13.483 & 3.04 & 79.88 & $-$229.42 & 58.71 & 5608 & 4.65 & $-$2.21 & 22:55:33.83 & $-$40:37:26.0 &  72.57 & $-$208.03 & 175.22 \cr 
\hline
 \label{tab_CPM_MP:tab_wide_comp_VHS}
 \end{tabular}
 \tablefoot{The primaries of the pairs are plotted as blue crosses in Fig.\ \ref{fig_CPM_MP:sample}. 
 The secondaries are not detected in $Gaia$. We give their coordinates and their proper motions (pmRA, pmDEC) computed from the cross-match between VHS and AllWISE\@.
 }
\end{table*}

\newpage
\section{FastCam data}
\label{CPM_MP:images_FastCam}
%

%
%
%
%
\begin{table*}
 \centering
 \caption[]{List of systems followed-up with lucky imaging. 
 }
\scriptsize
 \begin{tabular}{@{\hspace{0mm}}l @{\hspace{2mm}}c @{\hspace{2mm}}c @{\hspace{2mm}}c @{\hspace{2mm}}c @{\hspace{2mm}}c @{\hspace{2mm}}c @{\hspace{2mm}}c @{\hspace{2mm}}c @{\hspace{2mm}}c @{\hspace{2mm}}c @{\hspace{2mm}}c @{\hspace{2mm}}c@{\hspace{2mm}}c@{\hspace{2mm}}c@{\hspace{2mm}}c@{\hspace{0mm}}}
 \hline
 \hline
GaiaDR3\_ID         & R.A.        &  dec          &  Plx   &  PM    &  $G$  &  [Fe/H]  &  RUWE & $\epsilon_{i}$ & sig$_{\epsilon_{i}}$ & IPDfmp & Sep & Sep & Angle & $\Delta$Flux & $Gaia$ \cr
 \hline
                    &             &               &  mas   & mas/yr &  mag  &   dex  &       &        &         &      & pix & au  & deg &    &  \cr
 \hline
2863371034576236032 & 00:20:53.99 & $+$32:58:40.4 & 13.170 & 62.90  & 8.104 & $-$1.52 & 1.343 & 0.246  & 34.706  & 22   & 53.8 &  142.7 &  20.2 & 0.36 & Y$+$ \cr 
2806279443059663872 & 00:34:25.60 & $+$24:10:44.8 & 13.177 & 29.08  & 9.429 & $-$1.66 & 0.889 & 0.058  & 4.448  &  7   & 29.7 &   78.7 &  233.9 & 0.33 & Y  \cr
2530681282984436736 & 00:47:01.22 & $-$01:15:09.8 & 11.236 & 67.32  & 8.756 & $-$4.04 & 0.967 & 0.156  & 27.708  &  1   & 36.2 &  112.5 &  249.8 & 0.42 & Y$+$  \cr
522891484951192320  & 01:06:23.06 & $+$62:45:40.5 & 15.011 & 107.06 & 6.489 & $-$2.50 & 0.757 & 0.136  & 28.627  &  0   & 30.7 &  71.4 &  49.3 & 0.16 & Y?  \cr
132522762106356224  & 02:27:01.10 & $+$31:17:16.7 & 12.023 & 93.36  & 7.767 & $-$1.82 & 0.865 & 0.112  & 5.829  & 18   & 41.0 &  119.1 &  99.1 & 0.26 & Y  \cr
82603594175480320   & 02:28:22.97 & $+$17:21:57.9 & 10.566 & 96.67  & 9.131 & $-$3.98 & 1.847 & 0.261  & 40.854 & 21   & 34.4 &  113.7 &  302.5 & 0.76 & Y  \cr
541801332594262912  & 02:48:55.52 & $+$69:38:04.0 & 15.314 & 37.57  & 6.229 & $-$1.70 & 0.930 & 0.208  & 67.263  &  0   & 34.7 &   79.1 &  303.2 & 0.94 & Y$+$  \cr
231825330868662656  & 03:59:16.82 & $+$44:04:15.3 & 15.223 & 86.54  & 10.11 & $-$1.61 & 0.887 & 0.053  & 4.237  &  0   & --- & --- & --- & --- & N  \cr
3429156594127821440 & 05:50:47.77 & $+$26:03:15.0 & 10.081 & 33.36  & 7.568 & $-$2.09 & 12.48 & 2.752  & 6151.35 &  0   & --- & --- & --- & --- & N  \cr    
3341880316053633024 & 06:04:08.30 & $+$11:00:56.4 & 11.655 & 38.10  & 8.284 & $-$2.84 & 1.075 & 0.140  & 16.553  & 12   & 36.0 &  107.9 &  359.8 & 0.36 & Y  \cr
1806151193855154304 & 20:07:45.16 & $+$13:25:09.4 & 10.361 & 205.8  & 9.038 & $-$3.49 & 2.029 & 0.282  & 96.064 & 45   & 20.9 &  70.5 &  317.7 & 0.30 & Y?  \cr
1859330929163344000 & 20:45:40.19 & $+$30:43:13.3 & 16.072 & 15.48  & 8.989 & $-$1.81 & 2.993 & 0.418  & 215.612 & 15   & --- & --- & --- & --- & Y  \cr   
1787990525934491776 & 21:11:58.90 & $+$17:43:25.5 & 29.877 & 907.6  & 7.231 & $-$1.53 & 1.042 & 0.132  & 18.145  &  0   & --- & --- & --- & --- & Y?  \cr   
1846740215346263936 & 21:24:34.06 & $+$26:10:28.5 & 14.494 & 49.43  & 5.585 & $-$1.52 & 0.949 & 0.313  & 176.60  &  0   & --- & --- & --- & --- & N  \cr
2832463659640297472 & 23:07:28.84 & $+$21:08:02.5 & 24.461 & 119.2  & 5.910 & $-$1.56 & 1.474 & 0.307  & 122.099 &  0   & 38.3 & --- & 269.6 & 0.11 & N  \cr 
2651460165690234112 & 23:09:32.71 & $+$00:42:19.0 & 13.704 & 1314.  & 9.715 & $-$1.73 & 3.356 & 0.657  & 467.011 & 23   & --- & --- & --- & --- & N  \cr
1926630627143493760 & 23:42:14.37 & $+$44:45:19.1 & 11.898 & 60.36  & 6.942 & $-$1.66 & 2.169 & 0.433  & 234.260 &  0   & --- & --- & --- & --- & Y  \cr 
\hline
2881804450094712320 & 00:01:23.65 & +39:36:36.6 & 20.027 & 51.37 &  9.431 & $-$2.44 & 1.807 & 0.218 & 71.092 & 17 & 38.1 & 66.5 & 254.8 & 0.53 & Y$+$  \cr 
364785939116867072  & 00:56:45.41 & +38:29:58.2 & 26.704 & 154.97 &  3.843 & $-$1.59 & 4.024 & 2.989 & 17643.600 &  0 & 24.1 & 31.5 & 313.3 & 0.12 & N  \cr 
487090527352385408  & 03:49:36.56 & +63:17:48.1 & 13.949 & 62.54 &  5.782 & $-$1.80 & 8.312 & 2.349 & 8442.640 &  2 & --- & --- & --- & --- & N  \cr 
3312759303913978624 & 04:26:20.87 & +15:37:05.5 & 21.395 & 89.91 &  4.406 & $-$2.20 & 5.042 & 1.764 & 9171.680 &  0 & --- & --- & --- & --- & N  \cr 
274326712321580544  & 04:53:37.53 & +55:40:44.9 & 14.299 & 57.73 & 10.760 & $-$1.52 & 2.080 & 0.225 & 60.027 & 49 & 29.5 & 72.1 & 358.5 & 0.37 & Y  \cr 
267413670398436224  & 05:17:26.91 & +54:43:49.0 & 10.499 & 208.90 & 10.120 & $-$2.25 & 6.064 & 0.798 & 1263.540 &  0 & --- & --- & --- & --- & Y$+$  \cr 
3102275223166541184 & 06:47:59.34 & -04:15:06.2 & 10.565 & 30.54 &  7.609 & $-$1.80 & 9.729 & 2.283 & 3090.160 &  0 & --- & --- & --- & --- & N  \cr 
1084539101199508480 & 07:54:27.88 & +60:07:57.8 & 12.266 & 104.30 & 10.710 & $-$1.80 & 3.140 & 0.336 & 119.307 & 72 & 24.9 & 70.9 & 258.4 & 0.37 & Y?  \cr 
3066078265288036352 & 08:23:01.67 & -04:34:10.0 & 14.385 & 54.43 &  9.442 & $-$1.78 & 2.500 & 0.260 & 85.222 & 19 & 27.8 & 67.5 & 309.4 & 0.28 & Y  \cr 
1014058103758571520 & 08:59:11.75 & +48:02:27.1 & 68.000 & 489.10 &  3.120 & $-$1.75 & 4.840 & 1.637 & 2997.570 &  4 & 47.7 & 24.5 & 166.8 & 0.39 & N  \cr 
3762120497774235648 & 10:36:59.92 & -08:50:24.5 & 26.303 & 92.85 &  6.987 & $-$1.88 & 16.190 & 2.755 & 5067.550 &  6 & 27.9 & 37.1 & 313.0 & 0.20 & N  \cr 
3789997584304886912 & 11:16:39.58 & -03:39:06.4 & 17.282 & 115.80 &  4.417 & $-$2.16 & 2.430 & 1.174 & 1713.710 &  0 & --- & --- & --- & --- & N \cr 
860909637385712000  & 11:16:18.94 & +59:56:35.6 & 10.067 & 26.00 &  6.877 & $-$1.79 & 3.424 & 0.842 & 781.771 &  0 & --- & --- & --- & --- & N \cr 
1249265734250293376 & 13:40:40.78 & +20:00:49.5 & 10.607 & 49.15 &  7.320 & $-$1.62 & 2.404 & 0.557 & 231.446 &  0 & --- & --- & --- & ---  & N \cr 
3671065679270537344 & 14:10:55.48 & +04:24:16.7 & 16.052 & 140.10 &  8.368 & $-$1.78 & 8.504 & 1.156 & 1134.860 &  6 & 39.8 & 86.6 & 275.4 & 0.19 & Y  \cr 
1918049351204470016 & 23:13:39.06 & +39:24:57.5 & 11.886 & 360.80 & 10.790 & $-$1.95 & 6.744 & 0.901 & 1418.710 &  1 & --- & --- & --- & --- & N  \cr 
2419964589531462784 & 23:39:47.12 & -14:13:20.5 & 24.901 & 70.38 &  4.910 & $-$1.84 & 5.606 & 3.731 & 16840.200 &  0 & --- & --- & --- & --- & N  \cr 
132239603508139392  & 02:19:51.71 & +30:47:17.4 & 10.098 & 32.81 &  8.880 & $-$3.97 & 0.985 & 0.106 & 8.185 & 32 & 31.0 & 107.2 & 49.8 & 0.31 & Y?  \cr 
5184938090858917504 & 02:45:59.69 & -04:57:23.2 & 10.686 & 56.89 &  7.618 & $-$1.87 & 0.922 & 0.128 & 8.871 & 35 & 28.9 & 94.5 & 100.0 & 0.25 & Y?  \cr 
450345084978808704  & 03:35:00.96 & +60:02:29.6 & 19.799 & 34.38 &  6.721 & $-$4.10 & 0.916 & 0.132 & 20.133 &  4 & 47.5 & 83.8 & 0.6 & 0.34 & Y  \cr 
3311221224583178368 & 04:22:44.30 & +15:03:21.7 & 21.187 & 115.10 &  7.138 & $-$1.72 & 1.160 & 0.160 & 23.354 & 20 & 33.5 & 55.2 & 52.2 & 0.30 & Y  \cr 
271083015576372096  & 04:24:32.80 & +50:50:41.6 & 20.322 & 65.35 &  6.851 & $-$1.58 & 1.083 & 0.163 & 31.812 & 22 & 33.1 & 56.9 & 215.3 & 0.12 & Y  \cr 
3443553118342690048 & 05:47:24.93 & +29:39:24.2 & 13.161 & 123.40 &  8.076 & $-$1.83 & 1.049 & 0.152 & 12.984 & 26 & 50.0 & 132.7 & 214.3 & 0.54 & Y$+$  \cr 
1001751471970061184 & 06:46:14.09 & +59:26:29.9 & 13.310 & 20.30 &  5.493 & $-$2.10 & 1.323 & 0.838 & 1259.200 &  4 & 55.1 & 144.6 & 155.1 & 0.53 & Y$+$,Y  \cr 
713761132951480960  & 08:59:16.91 & +34:56:47.4 & 10.163 & 9.49 &  9.673 & $-$1.58 & 1.376 & 0.142 & 21.091 & 14 & 47.7 & 163.9 & 167.3 & 0.38 & Y$+$  \cr 
5741468163389212160 & 09:27:56.01 & -08:58:27.1 & 11.397 & 130.70 &  9.844 & $-$1.92 & 1.152 & 0.103 & 13.680 & 26 & 29.9 &  91.6 & 138.01 & 0.34 & Y  \cr 
634098255638793600  & 09:29:00.70 & +19:17:19.1 & 12.703 & 41.49 &  8.220 & $-$1.60 & 0.889 & 0.103 & 8.889 & 21 & 28.3 & 77.8 & 314.6 & 0.28 & Y  \cr 
3835575669874492928 & 10:16:54.84 & +02:25:16.9 & 13.062 & 187.60 &  8.723 & $-$1.81 & 1.247 & 0.152 & 16.286 & 23 & 25.4 & 67.9 & 0.9 & 0.22 & Y?  \cr 
731253779217024640  & 10:47:23.38 & +28:23:42.7 & 11.371 & 843.58 & 10.101 & $-$1.88 & 0.944 & 0.090 &  8.089  & 0  & --- & --- & --- & --- & N  \cr 
3789596743596878976 & 11:00:02.12 & -03:28:16.1 & 18.003 & 75.17  &  7.712 & $-$2.10 & 1.065 & 0.145 & 11.074 & 11 & 35.9 & 69.7 & 30.1 & 0.81 & Y  \cr 
3995455453806665728 & 11:12:30.34 & +24:39:07.3 & 12.577 & 49.82  &  8.041 & $-$1.52 & 1.015 & 0.264 & 41.018 &  0 & --- & --- & --- & --- & N  \cr
\hline
 \label{tab_CPM_MP:tab_LI_targets}
 \end{tabular}
 \tablefoot{The top part corresponds to observations on 18 October 2023\@. 
 while the bottom part to data collected on 14 December 2023\@. Objects without information on the separation and angle are classified as "single" based on the FastCam images presented in Appendix \ref{CPM_MP:images_FastCam}. The conversion from pixels to astronomical units (au) uses the plate scale derived for FastCam on the TCS (Sect.\ \ref{CPM_MP:search_LI}) and the $Gaia$ DR3 parallax. The last column indicates the presence of a companion with a $Gaia$ metallicity (Y$+$), without metallicity (Y), nearby object (Y?), and no companion (N), respectively.
 }
\end{table*}
\end{appendix}

\end{document}